\documentclass[10pt,preprint]{aastex}
\usepackage{natbib}

\newcommand{\fdot}{$\dot{f}$}
\newcommand{\rdot}{$\dot{r}$}
\newcommand{\ffdot}{$f$--$\dot{f}$}
\newcommand{\me}{\mathrm{e}}   
\newcommand{\mi}{\mathrm{i}}   
\newcommand{\sinc}{{\rm sinc}} 
\newcommand{\besi}{\mathrm{I}} 

\slugcomment{Submitted to The Astronomical Journal}

\shortauthors{Ransom, Eikenberry and Middleditch}
\shorttitle{Fourier Techniques}

\begin{document}

\title{Fourier Techniques for Very Long Astrophysical Time Series Analysis}

\author{Scott~M.~Ransom\altaffilmark{1,2},
  Stephen~S.~Eikenberry\altaffilmark{3}, and
  John~Middleditch\altaffilmark{4}}


\altaffiltext{1}{Department of Physics, McGill University, Montreal, QC 
  H3A~2T8; {\tt ransom@physics.mcgill.ca}}
\altaffiltext{2}{Center for Space Research, Massachusetts Institute 
  of Technology, Cambridge, MA 02139}
\altaffiltext{3}{Astronomy Department, 212 Space Sciences Building, 
  Cornell University, Ithaca, NY 14853}
\altaffiltext{4}{CCS-3, MS B256, Los Alamos National Laboratory, 
  Los Alamos, NM 87545}

\begin{abstract}
  
  In this paper, we present an assortment of both standard and
  advanced Fourier techniques that are useful in the analysis of
  astrophysical time series of very long duration~--- where the
  observation time is much greater than the time resolution of the
  individual data points.  We begin by reviewing the operational
  characteristics of Fourier transforms (FTs) of time series data,
  including power spectral statistics, discussing some of the
  differences between analyses of binned data, sampled data, and event
  data, and briefly discuss algorithms for calculating discrete
  Fourier transforms (DFTs) of very long time series.  We then discuss
  the response of DFTs to periodic signals, and present techniques to
  recover Fourier amplitude ``lost'' during simple traditional
  analyses if the periodicities change frequency during the
  observation.  These techniques include Fourier interpolation which
  allows us to correct the response for signals that occur between
  Fourier frequency bins.  We then present techniques for estimating
  additional signal properties such as the signal's centroid and
  duration in time, the first and second derivatives of the frequency,
  the pulsed fraction, and an overall estimate of the significance of
  a detection.  Finally, we present a recipe for a basic but thorough
  Fourier analysis of a time series for well-behaved pulsations.

\end{abstract}


\section{Introduction}

The analysis of time series data is an important tool in many areas of
astrophysics, including research involving white dwarfs, black holes,
and neutron stars.  In the study of neutron stars, time series
analysis has had particular importance for pulsar research due to the
high coherence of pulsar periodicities.  While many techniques can be
used to investigate the properties of these periodic signals, the
computational efficiency of the Fast Fourier Transform (FFT) makes
Fourier analysis the preferred approach for many applications
\citep[see e.g.][]{bc69}.  The literature contains literally hundreds
of descriptions of various aspects of Fourier analysis, most of which
deal with signal detection using the power spectrum.

Because of this concentration on power spectra, many researchers
discard a wealth of information provided by the Fourier phases.
Techniques which use this phase information exist that can provide
insight into many useful signal properties.  While many of these
techniques have been known for some time \citep[see e.g.][]{mid76},
few have appeared in textbooks or refereed journals, and fewer still
have been presented with any sort of derivation or insight into their
assumptions and/or limitations.

A second, more practical problem with most astronomical Fourier
analysis is its concentration on short time series.  We define
``short'' to mean either that the full time series of binned or
sampled data can fit into the core memory of your computer ($N
\lesssim 10^7$ points), or for data consisting of events (such as
photon arrival times in X-ray astronomy), that the time resolution
($dt$) of each event makes up a non-negligible fraction of the total
time duration ($T$) of the data ($T/{dt} \lesssim 10^7$).  Billion
point FFTs have been used successfully in the past
\citep[e.g.][]{and92, mkl+96, mkk+00}, but each required the use of
state-of-the-art supercomputing facilities.  Today, such analyses are
possible using clusters of workstations or even individual desktop
machines.  Many projects, such as pulsar searches of globular
clusters, astero- or helioseismological observations, and
gravitational wave experiments, require extremely large Fourier
transforms in order to make the highest sensitivity (i.e.\ 
\emph{coherent}) searches for pulsations, and to extract the maximum
amount of information from signals found in these searches.

It is our goal in this paper to present some useful Fourier analysis
techniques, that for various reasons are used only rarely when working
with long time series.  Most of our examples pertain to pulsar
searches of very long time series, but the methods can be used in the
Fourier analysis of virtually any coherent periodicity.  This paper
will briefly discuss the properties of the DFT, its response to
periodic signals and noise, and methods for its computation.  We will
discuss methods for interpolating Fourier amplitudes, estimating a
signal's duration and centroid in time, accurately determining its
frequency and frequency derivative, correcting for changing pulsation
frequency during an observation, and estimating the phase- or
amplitude modulation of a signal.  Such techniques allow the detection
of signals whose frequencies or amplitudes change significantly during
an observation --- for instance, due to orbital motion or pulsar
spin-down.

\section{Discrete Fourier Transform}

In order to present more advanced Fourier techniques later, we first
review some fundamentals of the DFT and its most common
implementation, the FFT.  Since thorough discussions of the Fourier
transform in both its continuous and discrete variants exist in the
literature \citep[e.g.][]{bra99}, we will mention only a few topics of
particular relevance to astrophysical data analysis, closely following
\citet{mid76}.

\subsection{Introduction to the DFT} \label{sec:dft}

The $k^{\rm th}$ element of the Discrete Fourier Transform (DFT), of a
uniformly-spaced time series $n_j$ ($j=0,1,\ldots,N-1$) is defined as
\begin{equation} 
  \label{ftdef}
  A_k = \sum_{j=0}^{N-1} n_j \ \me^{-2 \pi \mi jk/N},
\end{equation}
where $\mi = \sqrt{-1}$ and $k$ is the Fourier frequency or wavenumber
($k=0,1,\ldots,N-1$).  For a time spacing $dt$ between successive data
elements, the frequency of the $k^{\rm th}$ element is $f_k = k/(N\,dt)
= k/T$, where $T$ is the total time duration of the sequence being
transformed.  This frequency spacing of $1/T$, for evenly sampled,
un-padded and non-windowed data, is often called the Independent
Fourier Spacing (IFS)\footnote{The IFS is important when trying to
  determine the overall significance of a candidate from a search.
  The number of IFSs searched corresponds to the number of independent
  trials searched and should therefore be included in calculations
  that try to determine if a candidate was produced by noise.  See
  \citet{vvw+94}.} , and defines the finest frequency resolution
available while maintaining completely independent Fourier amplitudes.
Fourier frequency $N/2T$ is known as the Nyquist frequency.

If we view the DFT summation in the complex plane, we see that it is a
simple vector addition with each element rotated by $-2\pi k/N$ from
the previous element.  If the $n_j$ have a constant value, the sum
will form $k$ regular $N/k$-sided polygons with each polygon returning
near to the origin, and with the last one returning exactly to the
origin.  Therefore, the DFT of a constant data string will be
identically zero for all $k>0$, and equal to the sum of the data
elements for $k=0$ (the ``DC'' frequency element).

For most astrophysical observations, the data points are real-valued.
This property adds an important feature to Fourier transforms of these
time series --- they are symmetric about the Nyquist frequency, such
that $A_{N-k} = A_k^*$, where $A_k^*$ represents the complex conjugate
of $A_k$.  This symmetry allows us to calculate the full DFT of a time
series by computing amplitudes at only half of the normal Fourier
frequencies, thereby speeding up computation of the DFT by nearly a
factor of two \citep[e.g.,][]{ptvf92}.

When deriving properties and techniques based on DFTs, it is often
both computationally and intuitively easier to work with a
time-normalized data series, where $T=1$ and $u$ represents the
fraction of the observation complete at any given instant (such that
$0 \le u \le 1$).  In this case, instead of working with frequencies
$f$, or integral wavenumbers $k$, we define $r$, which represents any
real wavenumber (including non-integers).  If the number of samples,
$N$, from our data source, $n(u)$, is large, we can compute a
continuous FT
\begin{equation}
  \label{eq:ftintegral}
  A_{r}=N \int^{1}_{0}n(u)\ \me^{-2\pi \mi ru}\,\mathrm{d}u,
\end{equation}
which is almost identical to our DFT when $r=k$, and produces very
high accuracy estimates of the Fourier amplitudes at any frequency
such that $0 \ll r \ll N/2$.  We will use this approximation in many
of our derivations.

\subsection{Computation of Very Long FFTs} \label{sec:longFFTs}

The Fast Fourier Transform is a family of well-known computer
algorithms which quickly calculate a DFT in only $O(N\log_{2}(N))$
operations as opposed to the $O(N^{2})$ operations of a brute-force
DFT\@.  FFTs, their computation, and their origins have been described
in numerous articles and books over the last few decades
\citep[see][for an introduction]{bra99}.  Therefore, we will describe
only a few special versions which have become generally known only
recently and which are useful in the analysis of extremely long time
series.

High energy pulsar searches using photon counting systems (infrared,
optical, x-ray, and gamma-ray detectors) or pointed radio telescope
searches (for example, searching globular clusters) often utilize very
high sampling rates (i.e.\ $10-50$ kHz) and very long integration
times of hours, days, or even weeks.  These observations result in
time series with hundreds of millions or even billions of data points.
The subsequent FFTs are impossible to perform using conventional FFT
algorithms unless the full time series fits into the core memory of
the computer being used.  Utilizing special ``out-of-core'' FFT
methods, we can transform such huge arrays using distributed memory
parallel systems or with a single workstation and manageable numbers of
passes through the data stored on external media.  Most of these
methods are based on the ``four-step'' FFT and/or special external
media array permutation methods \citep[]{fra76, bai90a}.

\subsubsection{The ``Four-Step'' FFT}

Closely following the derivation described in \citet[]{sw95}, we can
think of our one-dimensional time series as a ``C-like'' or
row-ordered, two-dimensional matrix of size $N=N_rN_c$, where $N_c$ is
the number of columns (i.e.\ the length of a row) and $N_r$ the number
of rows (i.e.\ the length of a column).  Using this data
decomposition, the FFT is computed by \emph{1.}\ FFTing the columns of
the matrix, \emph{2.}\ multiplying the data by complex ``twiddle
factors'', \emph{3.}\ FFTing the rows of the matrix, and \emph{4.}\ 
matrix transposing the result.  If we define indices $x = 0,1, \ldots
,N_c-1$, $\ y = 0,1, \dots ,N_r-1$, $\ l = 0,1, \dots ,N_r-1$, and $m
= 0,1, \dots ,N_c-1$, we can write our signal and its DFT amplitudes
as
\begin{equation}
  n(x,y) = n_j, \; j = N_c \ y+x
\end{equation}
and
\begin{equation}
  A(l,m) = A_k, \; k = N_r \ m+l.
\end{equation}
Substituting into the definition of the DFT (eqn.~\ref{ftdef}) and
simplifying we get
\begin{equation}
  A(l,m) = \sum_{x=0}^{N_c-1} \left[
  \me^{-2 \pi \mi xl/N} \sum_{y=0}^{N_r-1} n(x,y) \me^{-2
      \pi \mi yl/N_r} \right] \ \me^{-2 \pi \mi xm/N_c}.
\end{equation}
Notice that the bracketed terms are really $N_c$ DFTs of length $N_r$
--- FFTs of the matrix columns multiplied by the ``twiddle factor''
$\me^{-2 \pi \mi xl/N}$. We denote these column FFTs as
\begin{equation} 
  A_c(x,l) = \me^{-2 \pi \mi xl/N}
  \sum_{y=0}^{N_r-1} n(x,y) \ \me^{-2 \pi \mi yl/N_r}.
\end{equation}
We now see that the outer summation is $N_r$ DFTs of length $N_c$ ---
FFTs of the matrix rows composed of the $A_c(x,l)$ terms.  We denote
the result of these transforms as
\begin{equation}
  A_r(m,l)
  = \sum_{x=0}^{N_c-1} A_c(x,l) \ \me^{-2 \pi \mi xm/N_c}.
\end{equation}
To recover the full FFT in its normal order, we simply transpose
$A_r(m,l)$
\begin{equation}
  A(l,m) = A_r^{T}(m,l).
\end{equation}

The ``four-step'' algorithm only needs small portions of the data in
memory at any one time.  Unfortunately, the short FFTs are strided in
memory\footnote{``Strided'' means that the data is stored in a
  sequence of non-contiguous memory locations spaced by a constant
  amount of memory known as the ``stride''.} by instead of being
stored contiguously.  This results in a significant inefficiency in
today's cache-based processors and requires inter-node communications
when distributing the short FFTs over many processors on parallel
computer systems.  These shortcomings can be overcome with the
``six-step'' algorithm \citep[]{bai90a}.

\subsubsection{The ``Six-Step'' FFT}

If the initial data set is transposed from a $N_c \times N_r$ matrix
into a $N_r \times N_c$ matrix, the strided column FFTs of length
$N_r$ become contiguous row FFTs.  This memory locality facilitates
the use of processor cache systems, greatly increasing memory response
times, and allows independent calculation of the row FFTs by different
processors in parallel systems.  A second transpose operation after
the row FFTs have been calculated counteracts the effects of the first
transpose operation and makes the next set of row FFTs contiguous in
memory as well.  The FFT is finished with a final transpose operation
which brings the data into normal order.

These transpose operations are relatively efficient on distributed
memory parallel systems with fast inter-node communications, and add
little time to the overall FFT.  In fact, for applications with very
large $N$, the time required to move the data from external media to
computer memory and vice versa tends to dominate the FFT time.  The
``six-step'' algorithm has become the ``standard'' FFT algorithm for
distributed memory systems where the serial nature and large number of
short FFTs exploit parallel computation strengths.

Unfortunately, if the data does not all fit into the core memory of a
workstation or parallel machine, the transposes become extremely slow
operations since data must be read to and written from much slower
external media.  \citet[]{fra76} devised optimized methods for dealing
with such data permutations on external media including the
``two-pass'' FFT algorithm.

\subsubsection{The ``Two-Pass'' Out-of-Core FFT}

\citet[]{fra76} and \citet[]{bai90a} both describe how a very large
data set may be Fourier transformed with only two read-write passes
through externally stored data if a scratch area on the external media
the same size as the input data set exists.  The method uses the
``four-step'' algorithm with out-of-core transposes.  These transpose
algorithms allow blocked external media access and perform most of the
transposition work in core memory.  While external media access speeds
and transfer rates are orders of magnitude slower than core-memory
systems, the ``two-pass'' algorithm allows huge arrays to be
transformed in manageable times.

\section{Fourier Transforms of Real Data}

\subsection{Fourier Response to Noise} \label{sec:noise}

If our data $n_j$ are composed of some constant value $c_j$ plus
random real-valued noise fluctuations $d_j$ with zero mean, the
transform terms become
\begin{equation} 
  (c_j + d_j) \ \me^{-2 \pi \mi jk/N} = 
  c_j \ \me^{-2 \pi \mi jk/N} + d_j \ \me^{-2 \pi \mi jk/N}.
\end{equation}
The linear nature of the Fourier transform allows us to treat the DFT
of the $d_j$ independently from the constant length steps, $c_j$.
Since the complex phase factor for a given $j$ and $k$ is fixed, the
direction of each element in the sum is nearly fixed.  However, since
the sign of the $d_j$ may be either positive or negative, the vector
direction of the $j^{\rm th}$ element may be reversed.  Thus, the DFT
of the $d_j$ creates a kind of random walk in the complex plane.

The statistical properties of this random walk for DFTs of pure noise
have been well studied \citep[see, e.g.][]{bt59}, and result in power
spectra distributed according to an exponential distribution (a
$\chi^2$ distribution with 2 degrees of freedom) with average and
standard deviation equal to $N\left< d_j^2 \right>$.  If we normalize
the powers by dividing by $N\left< d_j^2 \right>$, the probability for
a power $P = \left| A_k \right|^2$ in a single bin to equal or exceed
a power $P'$ by chance is\footnote{This is different than the
  probability for an actual signal to produce a power $P>P'$ in the
  presence of noise (see~\S\ref{sec:signoise}).  This difference is
  important in setting upper limits on the amplitudes of periodic
  signals as discussed in \citet{vvw+94}.}
\begin{equation}
  \label{powprob}
  Prob(P \geq P') = \me^{-P'}.
\end{equation}
Similarly, if we sum $m$ properly normalized powers, the probability
for the summed power $P_m$ to exceed a power $P'$ is given by
\begin{equation}
  \label{powsumprob}
  Prob(P_m \geq P') = \sum_{j=0}^{m-1} \frac{(P')^j}{j!} \ \me^{-P'}, 
\end{equation}
which is the probability for a $\chi^2$ distribution of $2m$ degrees
of freedom to exceed $2P'$.  Such an \emph{incoherent} (since no phase
information is used) summation of powers is often useful when
searching for signals suspected of having power in many harmonics (see
\S\ref{sec:nonsinusoid}).

Proper normalization of the powers is essential for an accurate
estimate of a signal's statistical significance or lack thereof.  We
often cannot normalize our power spectrum by simply dividing by
$N\left< d_j^2 \right>$, since frequency-dependent noise may be
present throughout our power spectrum --- perhaps as a result of
instrumental, atmospheric, or astrophysical processes.  Typically,
these processes produce noise which increases in strength towards the
low-frequency part of the spectrum and is correspondingly called
\emph{red noise}.

Techniques to flatten or remove this ``coloured'' noise component from
the power spectrum are described by \citet{is96}, and usually involve
dividing short portions of the power spectrum by the locally
determined average power level, $P_{local}$, such that
\begin{equation}
  \label{locpow}
  P_{k,norm} = {\frac{\left| A_k \right|^2}{N\left< d_j^2 \right>}} 
  \simeq {\frac{\left| A_k \right|^2}{P_{local}}} 
  = \frac{P_k}{P_{local}}.
\end{equation}
As long as the number of averaged powers is small enough such that the
power spectrum is roughly constant over the range in question, a
\emph{white noise} like power spectrum is produced with average and
standard deviation of approximately one and an exponential
distribution (eqn.~\ref{powprob}).

Since strong narrow-band signals near some frequency of interest will
skew a local power average upwards (and correspondingly decrease the
calculated significance of a signal detection), it is important to
exclude such powers from the calculation of $P_{local}$.  A simple and
effective way to accomplish this is by normalizing with a corrected
local median power level instead of the local power average.  An
exponential distribution with unity mean and standard deviation has a
median of $\ln(2)$.  Therefore, dividing a section of raw powers by
$1/\ln(2)$ times the local median value is theoretically equivalent to
normalizing with the local mean, but has the advantage of being
insensitive to high-power outliers in the spectrum.

More advanced algorithms for the removal of ``coloured'' noise and
power normalization do exist.  A simple example involves fitting
polynomial models to portions of the power spectrum and then dividing
by them out.  These methods work well for Fourier frequencies near
zero where the assumption of roughly equivalent power levels for the
local powers may be unwarranted.

For the special case where the noise in our data is purely Poissonian
(i.e.\ for binned photons in an optical or x-ray observation), we have
$\left< d_j^2 \right> = \left< n_j \right>$.  In this case, our
properly normalized power for the $k^{\rm th}$ DFT element is
\begin{equation}
  \label{pnorm}
  P_{k,norm} 
  = \frac{{\left| A_k \right|}^2}{N \left< d_j^2 \right>}
  = \frac{{\left| A_k \right|}^2}{n_{ph}},
\end{equation}
where $n_{ph} = N \left< n_j \right> $ is the sum of the $n_j$ (or
the total number of photons for a photon-counting system), which is
also equal to the ``DC'' frequency value of the FT (see
\S\ref{sec:dft}).  However, the same processes that caused the
``coloured'' noise discussed above can significantly alter this
situation and require a power normalization based on local powers (see
eqn.~\ref{locpow}).

\subsection{Fourier Response to Periodic Signals} \label{sec:signals}

One of the more useful properties of the FT for astronomical purposes
is its response to periodic signals.  Since all real periodic signals
can be expanded into a series of sinusoids it is important to
understand the FT response to a simple sine wave.

\subsubsection{Sinusoidal Signals} \label{sec:sinusoid}

If we now let our $n_j$ represent a sampled cosinusoid of amplitude
$a$, phase $\phi$, and frequency $f_r = r/T$ (where wavenumber $r$ is
an integer and $f_r$ an ``integral frequency''), we can write
\begin{mathletters}
  \begin{eqnarray}
    \label{eq:cosine}
    n_j 
    & = & a \ \cos (2 \pi f_r j\,dt + \phi) \\
    & = & a \ \cos (2 \pi j r/N + \phi) \\
    & = & \frac{a}{2} \left(\me^{2 \pi \mi jr/N + \mi \phi} + 
    \me^{-2 \pi \mi jr/N - \mi \phi}\right).\label{eq:cosineexp}
  \end{eqnarray}
\end{mathletters}%
From this expression, we see that the $k^{\rm th}$ element of the DFT
is given by
\begin{mathletters}
  \begin{eqnarray}
    A_k 
    & = & \frac{a}{2} \sum_{j=0}^{N-1} \me^{-2 \pi \mi jk/N} 
    \left(\me^{2 \pi \mi jr/N + \mi \phi} + 
      \me^{-2 \pi \mi jr/N - \mi \phi}\right)\\ 
    \label{eq:fftcosine}
    & = & \frac{a}{2} \sum_{j=0}^{N-1} \me^{2 \pi \mi j(k-r)/N + \mi
      \phi} + \me^{-2 \pi \mi j(k+r)/N - \mi \phi},
  \end{eqnarray}
\end{mathletters}%
and represents the summation of two vectors in the complex plane.  For
$k \neq r$, the first term traces out $\left| k-r \right|$ complete
``rotations'' (pseudo-polygons which start and end at the origin) in
the complex plane (since $k-r$ is an integer), giving a net
contribution of zero to the $k^{\rm th}$ DFT element. The second term
traces out $k+r$ complete rotations and once again contributes nothing
to the $k^{\rm th}$ DFT element (since $k$ and $r$ are both positive
and therefore $k+r \neq 0$.

When $k=r$, however, the consecutive terms in the summation add
coherently (i.e.\ in phase and therefore without rotation), since the
rotation caused by the DFT exponential exactly cancels that from the
signal exponential (the Fourier transform ``derotates'' the signal).
As a result, each element of the sum is a step in the complex plane of
magnitude $a/2$ in a direction parallel to the one set by the
arbitrary initial phase of the signal, $\phi$.  For a cosinusoidal
signal with integral frequency $f_r$, the DFT will be uniformly zero,
except in the $r^{\rm th}$ frequency bin, where the response is
\begin{equation}
  \label{eq:theocosresp}
  A_r = \frac{Na}{2} \ \me^{\mi \phi}.
\end{equation}

The Fourier response is more complicated for sinusoids with
non-integral frequencies (i.e.\ wavenumber $r$ is a non-negative real
number).  The $k^{\rm th}$ DFT element is still given by
eqn.~\ref{eq:fftcosine}, but not all of the signal ends up in a single
DFT bin $A_k$.  When $k=[r]$ (where $[r]$ is the nearest integer to
$r$), the first term in eqn.~\ref{eq:fftcosine} traces out a fraction
($k-r$) of a complete rotation in the complex plane, while the second
term traces out $k+[r]$ complete rotations, plus an additional
fractional rotation.

\begin{figure}
  \plotone{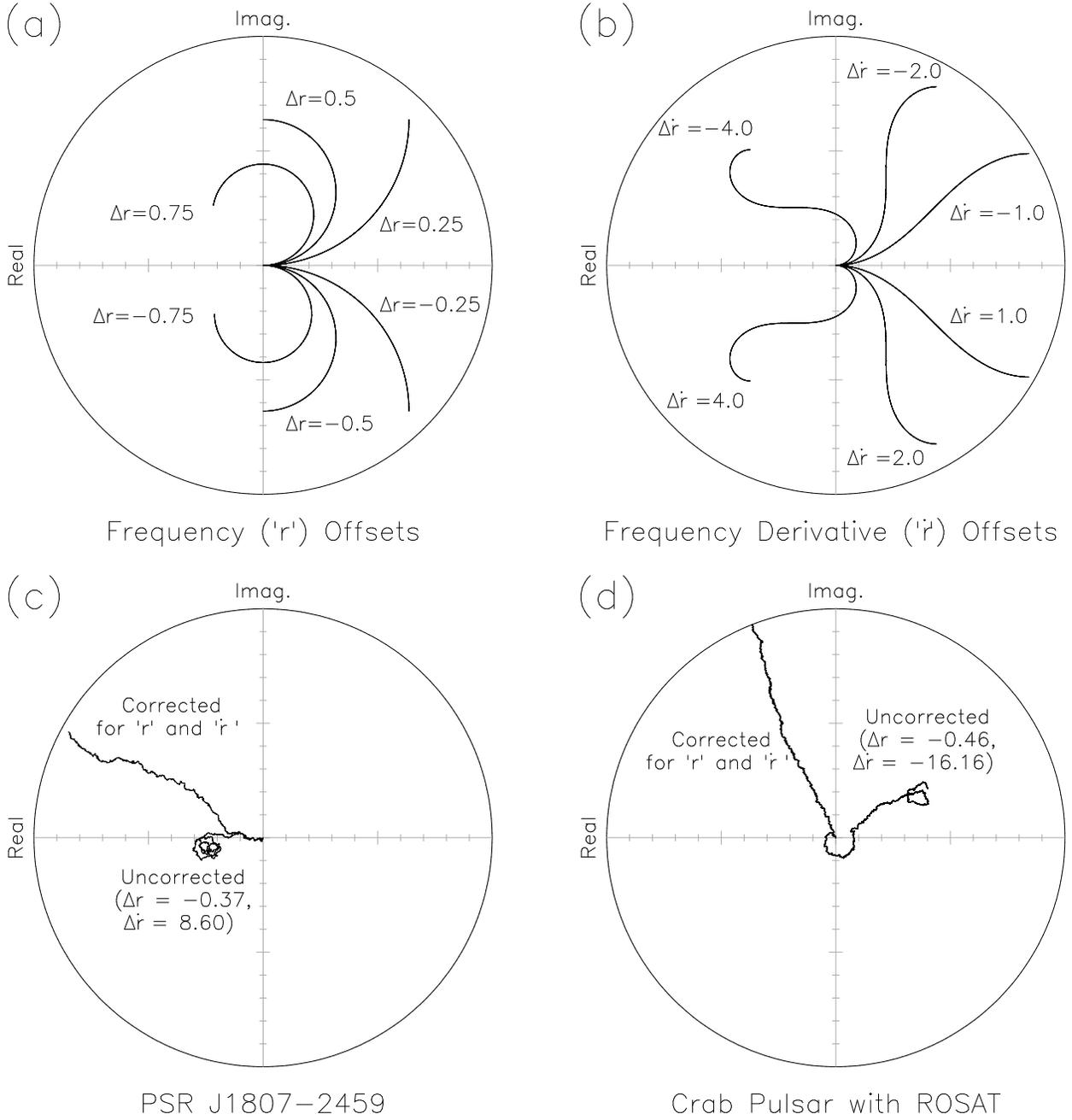}
  \caption{\footnotesize
    Fourier responses plotted as a series of vector additions in the
    complex plane.  The outer circles in each plot show the Fourier
    amplitude of a signal where all power is recovered by the vector
    addition (i.e.\ calculation of the DFT at the correct signal
    frequency $r$ and frequency derivative $\dot{r}$ for signals with
    linear changes in frequency over time).  The end-points of the
    vector additions are the Fourier amplitudes.  For plots (a) and
    (b), a fully recovered signal would start at $0+0i$ and end at
    $1+0i$.  (a) and (b) show the effects on Fourier amplitude and
    phase when a signal's intrinsic frequency ($r$ in bins or
    wavenumber) or frequency derivative ($\dot{r}$ in
    bins/observation) differs from the computed values.  For plot (b),
    the average Fourier frequencies in each case were correct, and
    only the frequency derivatives were in error.  (c) shows the
    response of PSR~J1807-2459 during its discovery observation
    \citep{rgh+01} with and without corrections for pulsar
    acceleration ($\dot{r}$) and interpolation in Fourier frequency
    ($r$).  The vectors were calculated using the method shown in
    \S\ref{sec:vector}.  The fact that the corrected vector does not
    quite reach the circle implies that higher order effects of the
    orbital motion remain un-corrected (see Fig.~\ref{fig:NGC6544}).
    (d) shows corrected and uncorrected responses of 10,000 randomly
    selected photons from a 2.4 day ROSAT observation of the Crab
    pulsar.
    \label{fig:vectors}}
\end{figure}

When $N$ is large, these complete and fractional ``rotations'' can be
treated as circles and arcs respectively.  Therefore, the first term
of eqn.~\ref{eq:fftcosine} results in a semi-circular arc of length
$Na/2$ {\it along the arc}, while the second term produces a
semi-circular arc of length $a/2\;(N\,\mathrm{mod}\,k+[r])$ along the
arc.  The DFT response is simply a vector drawn from the origin to the
end of the arc (see Fig.~\ref{fig:vectors}).  Since virtually all
astrophysical applications involve $r \gg 1$, where the first term
dominates the response, we will ignore the second term in the rest of
our analysis.

\placefigure{fig:vectors}

The final response is a chord subtending $2\pi(k-r)$ radians of a
circle of radius $Na/4\pi(k-r)$.  The equation for the length of a
chord is
\begin{equation}
  C = 2 \, \frac{s}{\Theta} \, \sin\left(\frac{\Theta}{2}\right),
\end{equation}
where $s$ is the arc length and $\Theta$ is the angle subtended by the
chord. The curvature of the arc away from the signal's starting phase
$\phi$ results in a phase change of $e^{-\mi \pi (k-r)}$.  Therefore,
the DFT response and power are
\begin{mathletters}
  \begin{eqnarray}
    A_k 
    & = & \me^{\mi \phi} \, \me^{-\mi \pi (k-r)} \, 2 \, 
    \frac{Na/2}{2 \pi (k-r)} \, \sin\left[\frac{2 \pi (k-r)}{2}\right] \\
    & = & \frac{Na}{2} \, \me^{\mi \phi} \, \, \me^{-\mi \pi (k-r)}
    \, \frac{\sin[\pi (k-r)]}{\pi (k-r)}  \\
    \label{eq:cosresp}
    & = & A_o \ \me^{-\mi \pi (k-r)} \, \sinc\left[\pi(k-r)\right] \\
    P_k 
    & = & \left| A_k \right|^2 = P_o \ \sinc^2 \left[\pi(k-r)\right],
  \end{eqnarray}
\end{mathletters}%
where $A_o = Na/2\;\me^{\mi \phi}$ is the DFT response for an integral
frequency signal (eqn.~\ref{eq:theocosresp}), $P_o$ is the
corresponding Fourier power, and the $\sinc$ function is defined as
$\sinc(x) = \sin(x)/x$.  This result is easily confirmed by a direct
integration of eqn.~\ref{eq:ftintegral} where $n(u)$ is equal to
eqn.~\ref{eq:cosineexp} with $j/N\to u$.

\begin{figure}
  \plotone{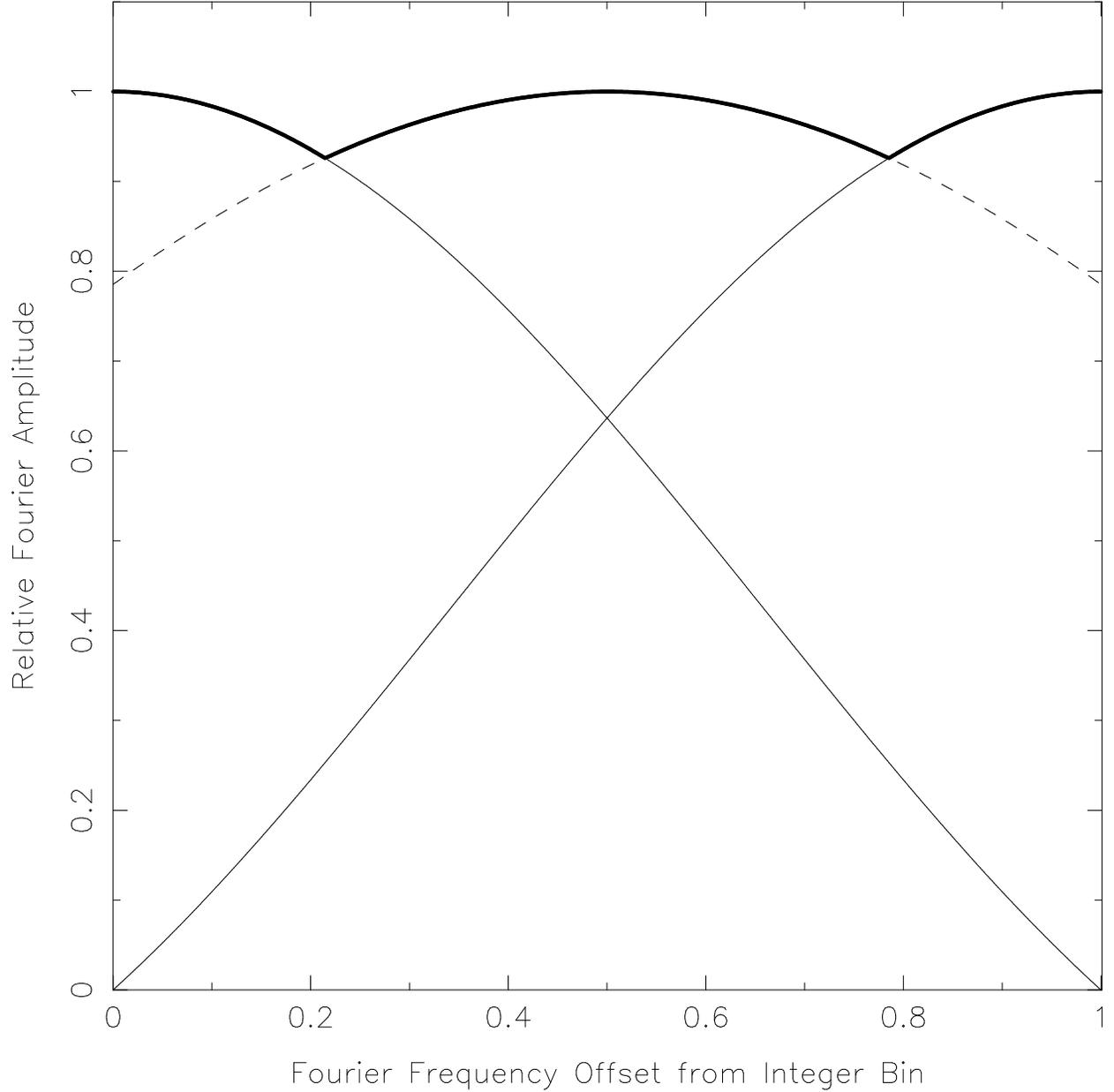}
  \caption{\footnotesize
    The thin solid black lines plot the $\sinc$ function amplitude
    responses of raw (or integer) FFT bins.  The point where these
    lines cross (at an offset of \onehalf\ and an amplitude of
    $1-2/\pi \sim 0.637$ of the full response) is the worst case
    response for an uncorrected FFT (see eqn.~\ref{eq:cosresp}).  The
    thin dashed line is the response of an ``interbin'' as calculated
    using eqn.~\ref{eq:interbin}.  The thick black line shows the
    overall ``scalloping'' when interbinning is used.  Worst case
    responses with interbinning occur at offsets of $\pm
    \left(1-\pi/4\right) \sim 0.215$ with amplitudes of $\sim 0.926$
    of the full response.
    \label{fig:interbin}}
\end{figure}

The $\sinc$ factor in eqn.~\ref{eq:cosresp} produces a loss of
sensitivity for the standard FFT to most real-world signals (where $r$
is not an integer).  This effect, often called ``scalloping''
\citep[e.g.][]{mdk93}, is shown in Fig.~\ref{fig:interbin}, and causes
a worst-case (when $\left| k-r \right| = 1/2$) amplitude reduction of
$\left| A_k \right| = 2\left| A_o \right|/\pi$ --- a nearly $60\%$
loss of signal power.  On average, scalloping results in a $\sim23\%$
loss of signal power \citep{van89}.  It is important to remember,
though, that this loss in sensitivity is due to the finite frequency
resolution of the FFT algorithm rather than an intrinsic feature of
the data itself.  In \S\ref{sec:interpolation} we discuss various
methods to reduce or even eliminate this loss of sensitivity.

\placefigure{fig:interbin}

\subsubsection{Non-Sinusoidal Signals} \label{sec:nonsinusoid}

Many real-world periodic signals are not sinusoidal.  Fortunately, we can
expand all real-valued pulsations as a series of $m$ sinusoidal
components
\begin{mathletters}
  \begin{eqnarray}
    \label{eq:cosseries}
    n_j 
    & = & \sum_{h=1}^{m} a_h \, \cos(2 \pi jhr/N + \phi_h) \\
    & = & \sum_{h=1}^{m} \frac{a_h}{2} 
    \left(\me^{2 \pi \mi jhr/N + \, \mi \phi_h} 
    + \me^{-2 \pi \mi jhr/N - \, \mi \phi_h}\right) 
  \end{eqnarray}
\end{mathletters}%
where $h=1,2, \ldots, m$ specifies the harmonic number (with $h=1$
known as the ``fundamental''), and $a_h$ and $\phi_h$ represent the
amplitude and phase of each component respectively.  Due to the linear
nature of the FT, we can treat the harmonics as independent sinusoidal
signals.  Each of these sinusoids produces a Fourier response
equivalent to eqn.~\ref{eq:cosresp}, except that $A_o$ becomes $A_h =
Na_h/2\;\me^{\mi \phi_h}$.

For nearly sinusoidal pulsations only the first few terms of
eqn.~\ref{eq:cosseries} contain significant amplitudes, $a_h$.  This
results in a similarly small number of significant peaks in the
corresponding power spectrum of the data.  Low duty-cycle pulsations
(i.e. those with a pulse that is short compared to the pulse period),
such as most radio pulsars, on the other hand, have dozens of
significant terms in their expansions and therefore harmonics in their
power spectra.

\begin{figure}
  \plotone{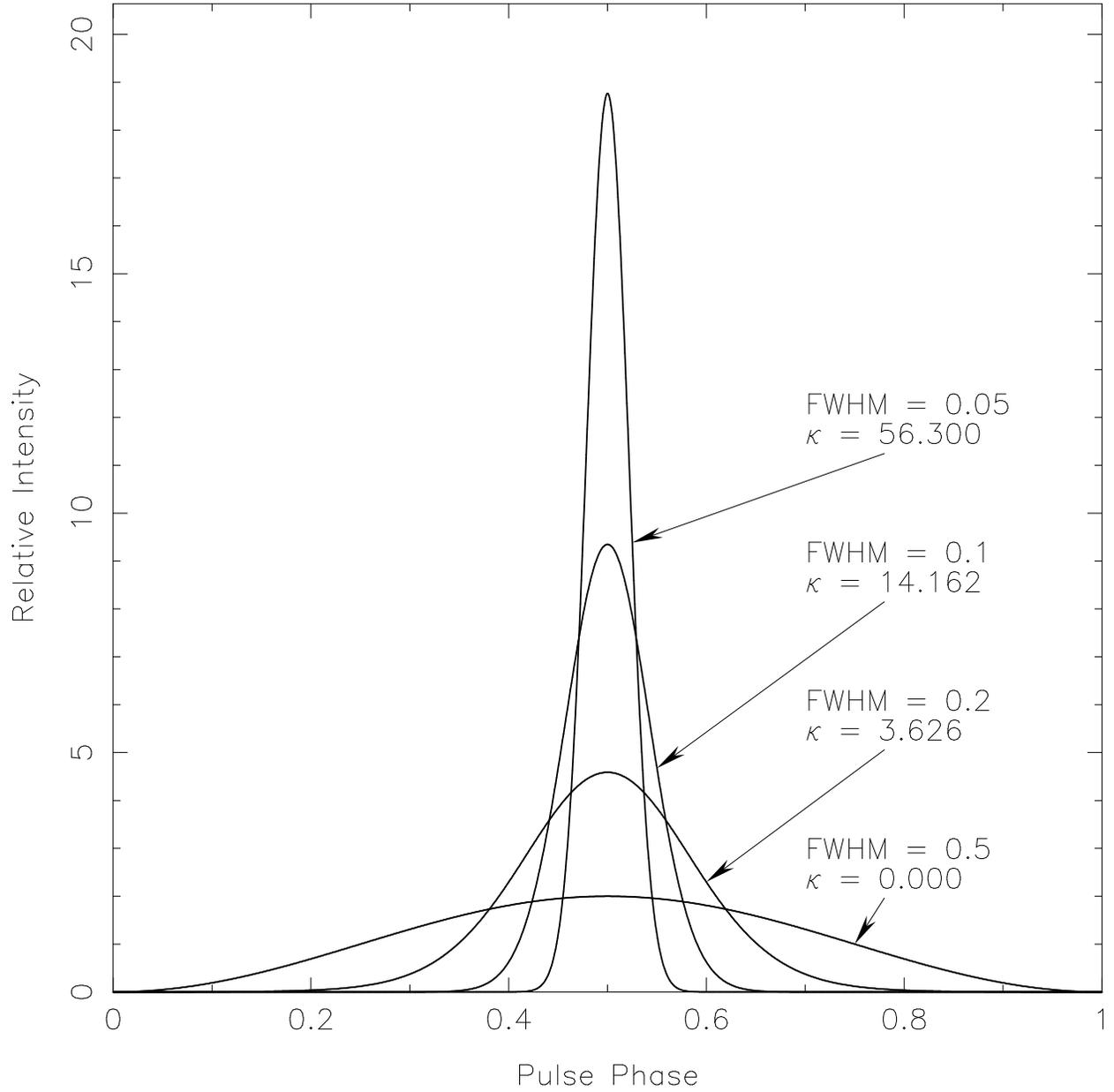}
  \caption{\footnotesize
    Sample pulse profiles from the modified von Mises distribution
    (MVMD) as described in \S\ref{sec:nonsinusoid}.  FWHM is the
    fractional full-width at half-maximum and $\kappa$ is the MVMD
    shape parameter.  High values of $\kappa$ result in Gaussian
    profiles, while as $\kappa \to 0$, the pulse shape becomes more
    and more sinusoidal.  The integral of a full pulse is equal to one
    unit, all of which is pulsed (i.e.\ the pulsed fraction is one).
    \label{fig:mvmd}}
\end{figure}

A useful pulsation model, particularly for radio and x-ray pulsars,
can be constructed based on a modified von Mises distribution (MVMD)
\begin{equation}
  \label{eq:mvmd}
  f(\kappa, t) = a \, \frac{\me^{\kappa\cos(2 \pi f_r t + \phi)} 
    - \me^{-\kappa}}{\besi_0(\kappa) - \me^{-\kappa}},
\end{equation}
where $0 \le t \le T$ is the instantaneous time, $\besi_0$ is the
modified Bessel function of zeroth order, and the shape parameter,
$\kappa$, determines the width of the function \citep[e.g.][]{mz75}.
In the limit as $\kappa \to 0$, the MVMD becomes a sinusoid, while as
$\kappa \to \infty$, it becomes a Gaussian (see
Fig.~\ref{fig:mvmd}).  The integral of the MVMD over a single pulse
period is simply $a$, all of which is pulsed (i.e.\ the pulsed
fraction is one).  The full-width at half-maximum (FWHM) as a
fraction of a pulse, is
\begin{equation}
  \label{eq:mvmd_ffwhm}
  {\rm FWHM_{MVMD}} = \pi^{-1}\arccos\left\{\ln\left[
        \cosh(\kappa)\right]\right\},
\end{equation}
and the maximum value is
\begin{equation}
  \label{eq:mvmd_max}
  {\rm max_{MVMD}} = \frac{2a\cosh(\kappa)}
  {\besi_0(\kappa) - \me^{-\kappa}}.
\end{equation}

\placefigure{fig:mvmd}

\begin{figure}
  \plotone{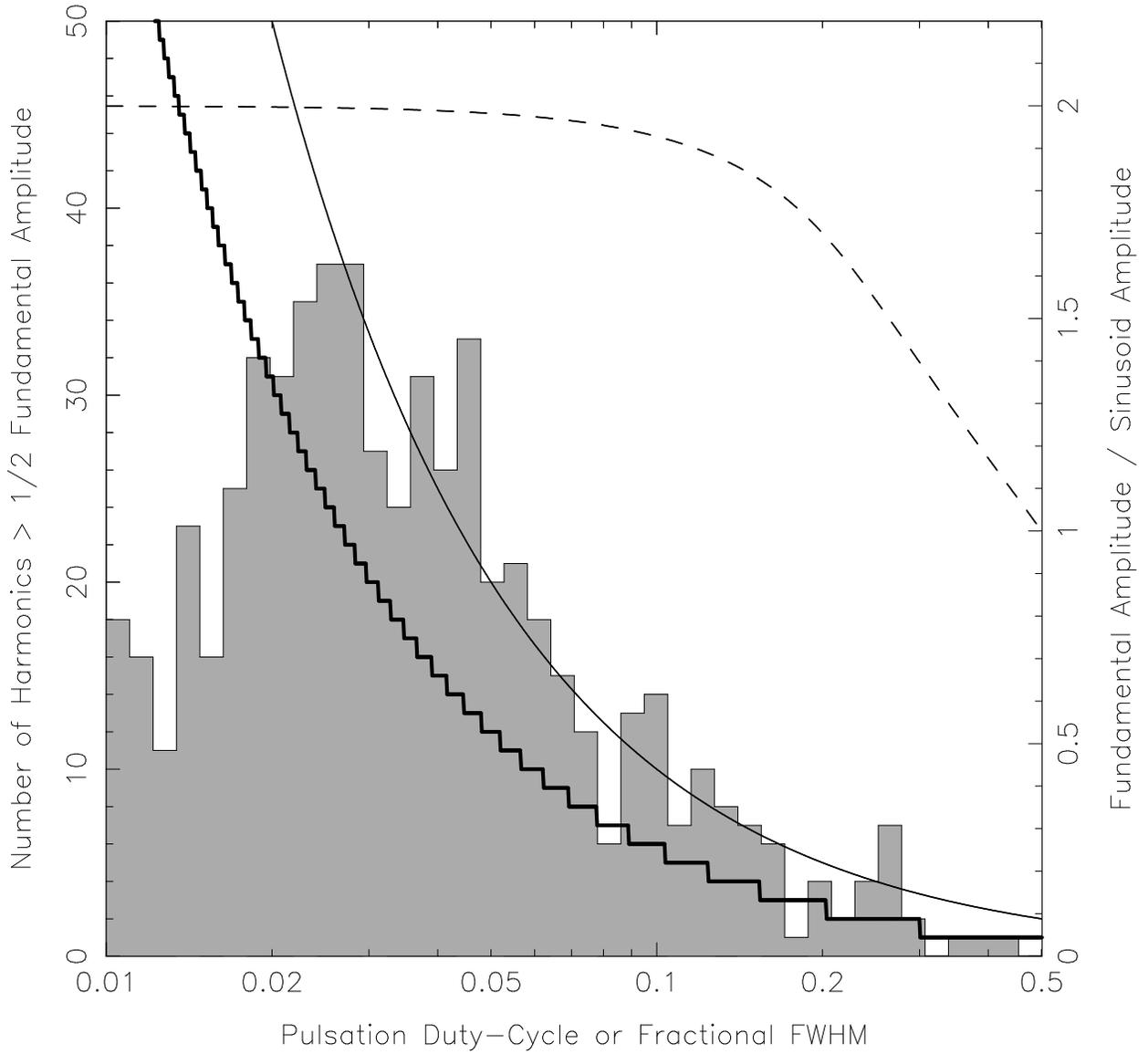}
  \caption{\footnotesize
    The thick solid line represents the approximate number of
    harmonics from an MVMD signal (see~\S\ref{sec:nonsinusoid} and
    Fig.~\ref{fig:mvmd}) that produce Fourier amplitudes greater than
    one-half the amplitude of the fundamental.  The thin solid line is
    the $1/{\rm FWHM}$ rule-of-thumb that is often used to
    estimate the number of significant harmonics a signal will
    generate.  The thin dashed line plots the ratio of the fundamental
    amplitude for an MVMD signal to a sinusoidal amplitude of the same
    pulsed intensity.  The grey histogram shows the distribution of
    pulse widths (FWHM) for over 600 radio pulsars.
    \label{fig:harmonics}}
\end{figure}

\begin{figure}
  \plotone{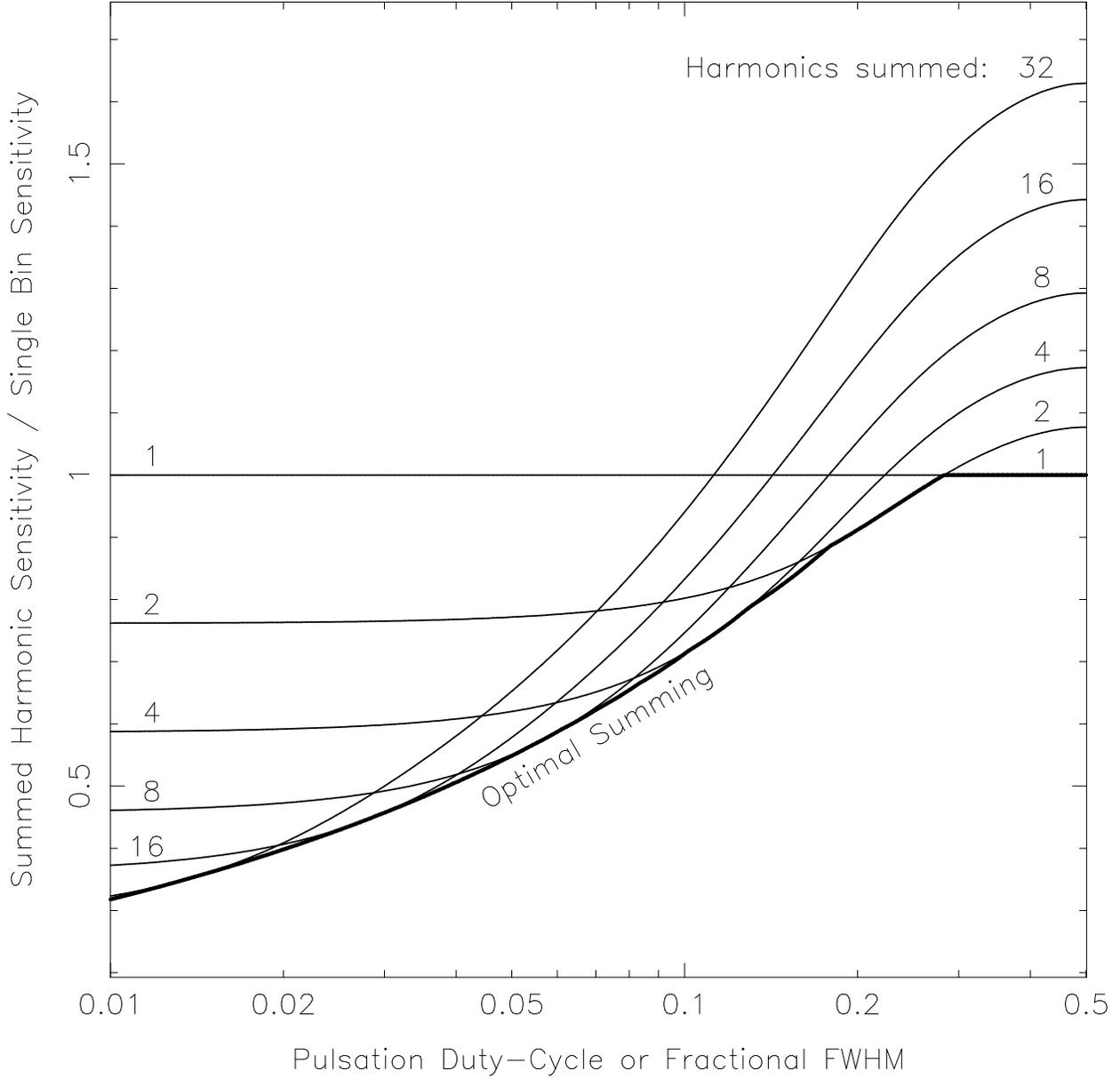}
  \caption{\footnotesize
    The thin lines represent sensitivities to MVMD signals (see
    \S\ref{sec:nonsinusoid} and Fig.~\ref{fig:mvmd}) for the
    incoherent summing of 1, 2, 4, 8, 16, or 32 harmonics, as compared
    to searches made without harmonic summing (\S\ref{sec:noise}).
    Lower numbers represent better sensitivity (i.e.\ fainter signals
    are detectable).  The best possible sensitivity using incoherent
    harmonic summing is shown by the thick black line.  It should be
    noted that incoherent summing produces worse sensitivities than
    not summing if the duty cycles of the pulsations are large.  This
    results from the fact that such pulsations have only a small
    number of significant harmonics (see Fig.~\ref{fig:harmonics}), so
    that summing tends to add only noise rather than signal.
    \label{fig:sensitivities}}
\end{figure}

The FT of the MVMD can be computed in a particularly convenient form
for harmonic analysis.  According to \citet[][eqn.~9.6.34]{as72}, we
can expand the exponential in the MVMD as
\begin{equation}
  \label{eq:expcos}
  \me^{\kappa\cos(x)} = \besi_0(\kappa) + 
  2\sum_{h=1}^{\infty}\besi_h(\kappa)\cos(hx),
\end{equation}
where $\besi_h$ is the modified Bessel function of order $h$.  When
combined with the rest of the MVMD definition we have
\begin{equation}
  \label{eq:mvmd2}
  f(\kappa, t) = a + \frac{2a\sum_{h=1}^{\infty}
      \besi_h(\kappa)\cos(2 \pi h f_r t + h \phi)}
    {\besi_0(\kappa) - \me^{-\kappa}}.
\end{equation}
This expression is simply a ``DC'' term (since the integral over a
pulse equals $a$) plus a series of independent cosinusoidal harmonics.
After Fourier transforming (i.e.\ substituting into
eqn.~\ref{eq:ftintegral} with $f_{r}t \to ru$), we are left with a
series of harmonics of amplitude
$AN\,\besi_h(\kappa)/\left[\besi_0(\kappa) - \me^{-\kappa}\right]$ and
phase $h\phi$ at Fourier frequency $hr$.  It is important to note that
as $\kappa\to\infty$ and the pulse becomes narrower, the Fourier
amplitudes of the low order harmonics are twice that of a sinusoid
with the same pulsed fraction (see the dashed line in
Fig.~\ref{fig:harmonics}).  This fact, along with the large number of
harmonics that low duty-cycle pulsations generate, can significantly
increase search sensitivities to such pulsations (see
Fig.~\ref{fig:sensitivities}).

\placefigure{fig:harmonics}

Fig.~\ref{fig:harmonics} shows the approximate number of significant
harmonics (meaning that a harmonic's amplitude is greater than
one-half the amplitude of the fundamental) generated by an MVMD
pulsation, as well as a histogram of the duty-cycles of over 600 radio
pulsars \citep[the majority of which are from][]{tmlc95}.  Most radio
pulsars have duty-cycles $\lesssim 5\%$ corresponding to $\gtrsim 10$
significant harmonics --- assuming a sufficient data sample rate.

\placefigure{fig:sensitivities}

\subsection{Periodic Signals With Noise} \label{sec:signoise}

When a periodic signal is present in a noisy time series, a sum of $m$
powers $P_m$, containing some amount of signal power $P_s$, is no
longer described by a $\chi^2$ distribution with $2m$ degrees of
freedom (see~\S\ref{sec:noise}).  \citet{gro75d} calculated the
expectation value and variance of $P_m$ as $\left< P_m \right> = m +
P_s$ and $\left< P_m^2 - \left<P_m\right>^2 \right> = m + 2P_s$
respectfully.  He also derived the exact probability density function
for $P_m$ which can be integrated to give the probability that $P_m$
is greater than or equal to some power $P'$,
\begin{equation}
  \label{eq:probsignoise}
  Prob((P_m;P_s) \geq P') = \me^{-\left(P'+P_s\right)}
  \sum_{k=0}^{\infty} \sum_{j=0}^{k+m-1} \frac{(P')^jP_s^k}{j!k!}. 
\end{equation}
When $P_s=0$ this equation reduces to Eqn.~\ref{powsumprob}.

The fact that the probability density function for a signal plus noise
is different from a $\chi^2$ distribution with $2m$ degrees of freedom
is very important when trying to determine the sensitivity of a search
for pulsations or an upper limit to the amplitude of a periodic signal
present in a time series.  \citet{vvw+94} describe a
procedure\footnote{Note that \citet{vvw+94} use a power normalization
  that is a factor of two higher than ours.} for correctly determining
search sensitivities and upper limits using the equations of
\citet{gro75d}.

\subsection{Photon Counting Data} \label{sec:photon}

Since many of today's astronomical time series come from photon
counting experiments, it is important to raise some of the issues
particular to Fourier analysis of such data.  If we can assume purely
Poissonian statistics, a power spectrum of pure noise is flat, and can
be normalized simply by dividing by the total number of photons in the
data (the zeroth or ``DC'' frequency bin from the DFT --- see
\S\ref{sec:noise}).  In addition to this difference in power spectrum
normalization, the other points worth noting come from the fact
that photon counting data is based on the measurement of events rather
than the instantaneous sampling of a continuous process.

One important issue that is beyond the scope of this paper is
dead-time correction.  Dead-time effects modify a detector's
sensitivity to photons for some time after the detection of an earlier
photon.  These effects can cause complicated non-linear and
frequency-dependent systematics during Fourier analysis.  We refer the
reader to \citet{zjs+95} and references therein, for a thorough
discussion of this topic.

\subsubsection{Binned vs. Sampled Data} \label{sec:binned}

Many high-energy telescopes and detectors produce time series of
binned photons rather than the sampled data produced by radio
telescopes.  Since binning essentially averages a periodic signal's
instantaneous rate over the binning time ($dt$), it modifies the
Fourier response to the signal.  Binning removes phase information
from the data and causes the Fourier response to sinusoidal pulsations
to become frequency-dependent --- resulting in decreased sensitivity
at high frequencies \citep[e.g.][]{mid76, lde+83}.

The frequency-dependent loss in Fourier amplitude due to binning is
$\sinc(\pi f_r dt)$ or $\sinc(\pi r/N)$.  The binned data Fourier
response to a sinusoid is therefore eqn.~\ref{eq:cosresp} times this
factor.  This decrease in sensitivity corresponds to a loss in signal
power of about $\sim19.8\%$ at half the Nyquist frequency and
$\sim59.5\%$ at the Nyquist frequency itself.

For Poissonian noise (i.e.\ from a photon-counting experiment that
does not introduce count-rate dependent systematics), which is
independent of the sampling interval, the Fourier response is flat
over all frequencies.  This is in contrast to a sinusoidal signal
passing through the same system which suffers the frequency-dependent
attenuation described above\citep{mid76}.  Such behavior is important
when trying to estimate limits or amplitudes for pulsations in a time
series \citep{vvw+94}.

\subsubsection{Low Count Rate Data} \label{sec:lowcount}

The Fourier analysis of gamma-ray or x-ray observations often places
us in a very unique regime --- very long integration times
($\gtrsim10^4$\,s) with very low numbers of counts ($\lesssim10^3$
photons).  In addition, due to visibility constraints based on the
orbits of the telescopes, large fractions of the time between the
first and last photons may be devoid of counts.

Fourier analysis of such data can overwhelm present computational
resources.  For example, a $10^6$\,s observation (about 11.6~days)
with photon time-of-arrival (TOA) resolution of $10^{-4}$\,s would
require a 10~Gpt FFT for a full-resolution analysis.  Such FFTs, while
possible, are extremely difficult to compute unless very special and
dedicated hardware resources are available.  If this data contains
only a small number of photons, however, we can exactly compute the
DFT over any frequency range and to any frequency resolution using a
brute-force implementation of the FT.

If we treat each TOA as a sample of amplitude one, an exact DFT
amplitude at arbitrary Fourier frequency $r$ becomes
\begin{equation} 
  \label{eq:dftexact}
  A_r = \sum_{j=1}^{n_{ph}} \me^{-2 \pi \mi r(t_j - t_o)/T},
\end{equation}
where $n_{ph}$ is the number of photons, $t_j$ is the TOA of the
$j^{th}$ photon, $t_o$ is the time of the start of the observation,
and $T$ is the total duration of the observation.  Very quick
harmonic-summing searches of an observation are possible using this
technique, with the added benefit that ``scalloping'' (see
\S\ref{sec:sinusoid}) is non-existent.

Since eqn.~\ref{eq:dftexact} only involves a summation over the number
of photons, it can be computed quickly if $n_{ph}$ is relatively
small.  Great increases in computation speed can be had if we search a
regular grid of Fourier frequencies.  Trigonometric recurrences such
as
\begin{mathletters}
  \begin{eqnarray}
    \label{eq:trigrecur}
    \cos(\theta + \delta) & = & \cos(\theta) -
    \left[\alpha\cos(\theta)+\beta\sin(\theta)\right] \\
    \sin(\theta + \delta) & = & \sin(\theta) -
    \left[\alpha\sin(\theta)-\beta\cos(\theta)\right],
  \end{eqnarray}
\end{mathletters}%
where $\alpha = 2\sin^2\left(\delta/2\right)$ and $\beta =
\sin(\delta)$, allow extremely efficient calculation of the complex
exponential for each TOA \citep{ptvf92}.  This technique allows one to
calculate billions of Fourier frequencies from a few hundred photons
using only modest computational resources.

\section{Improving the DFT Response to Arbitrary Signals}

\subsection{Fourier Interpolation} \label{sec:interpolation}

The potential for up to a 30\% decrease in signal-to-noise ($S/N
\propto \sqrt{P}$) due to an essentially arbitrary difference between
the signal frequency and the integer frequency of the nearest Fourier
bin is clearly a drawback in the use of the DFT (see
\S\ref{sec:sinusoid}).  However, if we could calculate the Fourier
Transform (FT) at a higher frequency resolution than the $1/T$ spacing
that results from the FFT, we could significantly reduce or eliminate
scalloping and effectively flatten the Fourier response as a function
of frequency.

One possibility for increasing the frequency resolution is to simply
calculate the DFT by a brute force summation at frequencies between
the integer frequencies.  Such a technique is possible in special
situations (see~\S\ref{sec:lowcount}), but for most applications, the
computational costs would be unacceptably high.  Another well-known
possibility is to ``pad'' the end of the time series with a large
number of points with values equivalent to the mean of the
data\footnote{Padding with the data mean is preferable to zero-padding
  since zero-padding introduces low frequency power into the Fourier
  response.}.  The padding adds no power to the data but it does
increase the Fourier resolution since $T$ has been artificially
increased by the padding.  While this technique is simple and
effective for short time series, the difficulties involved in
performing very long FFTs (\S\ref{sec:longFFTs}) makes this technique
difficult when dealing with long time series.

Yet another way to calculate a higher-resolution Fourier response is
to use the complex amplitudes produced by the standard FFT to
interpolate responses at non-integer frequencies --- a process known
as ``fine-binning'' or ``Fourier interpolation''
\citep[e.g.][]{mdk93}.  Similar techniques allow the full recovery of
a signal's theoretical coherent response provided that the signal's
behavior during the observation is either known or can be guessed.

The purpose of Fourier interpolation is to calculate a complex Fourier
amplitude at an arbitrary frequency $f_r=r/T$, where $r$ is any real
number, such that the result is sufficiently close to the exact
calculation,
\begin{equation}
  A_r = \sum_{j=0}^{N-1}n_j \me^{-2 \pi \mi jr/N}.
\end{equation}
We can rewrite this expression as
\begin{equation}
  \label{eq:conv1}
  A_r = \sum_{k=0}^{N-1}A_k\ \me^{-\mi\pi(r-k)} 
  \sinc\left[\pi(r-k)\right].
\end{equation}
where the $A_k$ are the complex FFT amplitudes at the integer
frequencies $l$ (see Appendix~\ref{app:fourinterp} and
\S\ref{sec:correlation} for a derivation and discussion of this
result).

The $\sinc$ function in eqn.~\ref{eq:conv1}, provides the key to
computing an accurate interpolated amplitude using a relatively small
number of operations.  Since $\sinc[\pi(r-k)]\to0$ as
$\pi(r-k)\to\pm\infty$, the expansion of $A_r$ in terms of the $A_k$
is dominated by the local Fourier amplitudes (i.e.\ where $k \sim r$).
We can therefore approximate $A_r$ as
\begin{equation}
  \label{eq:interp}
  A_r \simeq \sum_{k=[r]-m/2}^{[r]+m/2} A_k \ \me^{-\mi \pi(r-k)} 
  \sinc\left[\pi(r-k)\right], 
\end{equation} 
where $[r]$ is the nearest integer to $r$, and $m$ is the number of
neighboring FFT bins used in the interpolation.  Note that the
interpolation is simply a correlation of the local FFT spectrum around
the desired frequency element with a ``template'' response --- in this
case the theoretical response of a DFT to a sinusoid as described by
eqn.~\ref{eq:cosresp}.

\begin{figure}
  \plotone{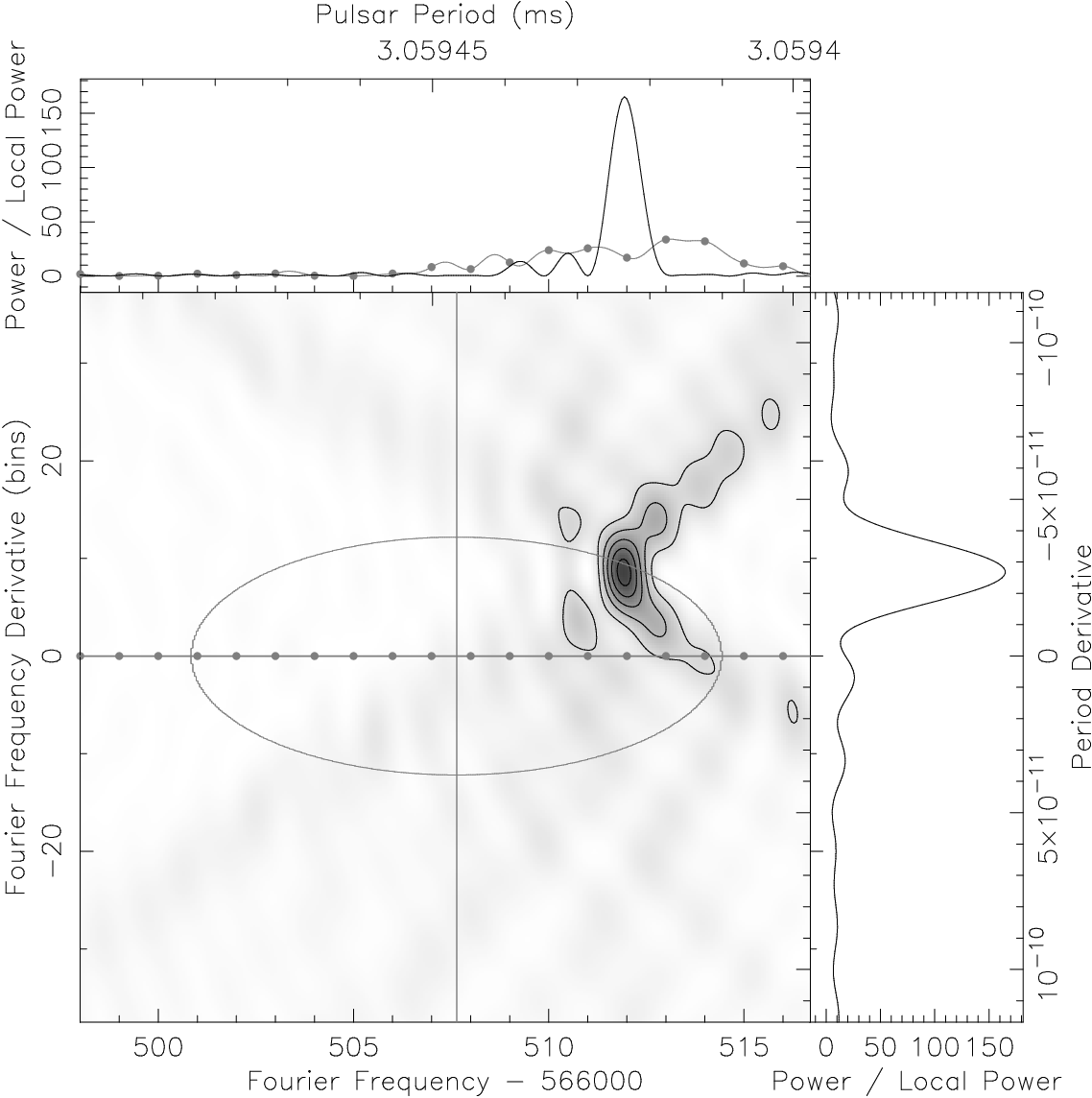}
  \caption{\footnotesize
    An $18\,\sigma$ (single trial) detection from the discovery
    observation of the 1.7\,hour binary PSR~J1807$-$2459 in the
    globular cluster NGC6544 using a Fourier-domain ``acceleration''
    search.  Contour intervals correspond to 30, 60, 90, 120, and 150
    times the average local power level.  The intrinsic pulsar period
    and $\dot{f} = 0$ (which corresponds to an un-accelerated FFT of
    the data) are marked by the solid gray lines.  The dots correspond
    to the ``raw'' or un-interpolated powers from the original FFT of
    the observation.  The gray ellipse is the predicted ``path'' of
    the pulsar in the \ffdot\ plane given the known binary parameters.
    During the 28.9\,minute observation, the pulsar moved from
    $\sim11$ o'clock to $\sim3$ o'clock on the ellipse.  The peak's
    slight offset from the ellipse and the presence of ``shoulders''
    indicate that the constant \fdot\ assumption of the acceleration
    search could not fully correct for the orbital motion during this
    observation.  The top and right-hand panels show cuts through the
    peak in the $f$ and \fdot\ directions respectively.  The line in
    the top panel with gray dots shows the Fourier interpolated
    $\dot{f}=0$ power spectrum (calculated as per
    \S\ref{sec:interpolation}).
    \label{fig:NGC6544}}
\end{figure}

The upper panel in Fig.~\ref{fig:NGC6544} shows both the raw FFT power
spectrum (denoted by grey dots) and the interpolated power spectrum
(the line connecting the dots) for a radio observation of the
short-period binary pulsar PSR~J1807$-$2459.  The spectra cover a
narrow frequency range near the pulsar's rotational frequency and were
calculated using $m=32$ and a frequency step size of $\Delta r = 1/16$
(compared to the raw FFT frequency step size of $\Delta r = 1$).  Note
that the interpolated spectrum reveals much more information regarding
the shape of the frequency profile --- including the true amplitude
and location of the maximum in Fourier power.  We will see in
\S\ref{sec:sigprops} how this can be used to deduce further
information regarding the signal properties.

\placefigure{fig:NGC6544}

A computationally less expensive version of Fourier interpolation is
``interbinning,'' where we approximate the FT response at half-integer
frequencies using only the nearest 2 integer frequency bins.  By using
the Fourier interpolation equation (eqn.~\ref{eq:interp}) with $m=2$,
ignoring an overall phase shift, and boosting the response such that
the best-case response (at half-integer frequencies) is equivalent to
the full response, we get
\begin{equation}
  \label{eq:interbin}
  A_{k+\frac{1}{2}} \simeq \frac{\pi}{4} \ (A_k - A_{k+1}).  
\end{equation}
This particular formulation of interbinning was reported by
\citet{van89}, and its response is shown in Fig.~\ref{fig:interbin}.
\citet{mdk93} have contributed a correction to this formula for use
when the data are padded at the end.

Interbinning is extremely useful since such a computationally
inexpensive calculation reduces the maximum loss of signal-to-noise
from $1-2/\pi$ or $\sim36\%$ at a frequency offset of \onehalf\ bin to
$\sim 7.4\%$ at an offset of $\pm\left(1-\pi/4\right)$ bins.  This
large but cheaply-obtained reduction in scalloping can be extremely
beneficial when searching large numbers of FFT bins and interbins.

It is important to note that interbins as defined above have three
different properties than integer FFT bins.  First, they have
different noise properties, which makes calculation of the
significance of interbin powers much more difficult.  Second, each
interbin is correlated with the integer bins it was created from,
meaning that interbins are not independent Fourier trials (see
\S\ref{sec:dft} for a discussion of the IFS).  And finally, interbins
do not recover the correct phase of a sinusoid at the interbin
frequency.  In general, since interbins are most commonly used during
searches to simply identify signals in the power spectrum that would
otherwise have been lost due to scalloping, these weaknesses do not
degrade the usefulness of their calculation.  When a signal is
identified, a full-scale interpolation of the Fourier amplitudes
around the signal using eqn.~\ref{eq:interp} allows accurate estimates
to be made of the signal's significance and other properties (see
\S\ref{sec:sigprops}).

\subsection{General DFT Response Correction} \label{sec:respcorr}

Fourier interpolation serves as a specific example of a much more
general technique --- the ability to completely recover the fully
coherent response for virtually any signal.  For Fourier
interpolation, we can exactly calculate the response of any Fourier
frequency based purely on the properties of the FT.  To correct the
Fourier transform's response to a particular signal, we must know not
only the properties of the FT, but the properties of the signal we are
looking for as well.  For the cases we will discuss, this ability
comes in one of two forms: matched filtering in the Fourier domain
using only the ``local'' Fourier amplitudes near the Fourier
frequencies of interest (which we call the ``correlation technique''),
or the straightening of the curved Fourier vector addition in the
complex plane (which we will call ``vector bending'').

\subsubsection{Correcting for Constant Frequency Derivative}

In order to illustrate these two methods, we demonstrate how to
correct for a signal whose response is reduced due to a constant
frequency derivative \fdot\ (or in Fourier frequency bins,
$\dot{r}=\dot{f}T^2$).  The DFT operates, as we noted in
\S\ref{sec:sinusoid}, by ``derotating'' the vector addition of the
data in the complex plane by changing the phases of each of the vector
elements --- causing a straight line to form for a sinusoidal signal
with integral frequency.  In the presence of a frequency derivative
however, the signal frequency may change by one or more frequency bins
over the course of the observation.  The complex phase corrections
provided by the DFT will fail to completely derotate the data, and
pulsation power will be ``smeared'' across several nearby frequency
bins --- causing a decrease in the measured DFT response
\citep[e.g.][]{jk91}.  Fig.~\ref{fig:vectors} illustrates this
effect in the complex plane.

As with a frequency error, an uncorrected frequency derivative causes
the vector addition to form an arc, although in this case
quasi-parabolic rather than circular.  The decreased DFT response
equals the distance from the origin to the end of the arc.  This
distance is significantly shorter than that of a coherently detected
signal which equals the the length along the arc.

Signals with constant or nearly constant \fdot\ are quite common in
pulsar astronomy --- especially when dealing with time series of very
long duration.  In such long observations, even the very small
spin-downs typical of pulsars can cause a signal to drift across
numerous Fourier bins.  The Doppler effects of binary pulsar orbits
cause similar frequency drifting when the observation time is much
shorter than the orbital period.

The ``standard'' method to correct for a constant frequency derivative
is to ``stretch'' the original time series to compensate for the known
or assumed \fdot.  This process involves re-sampling the data ensemble
$n_j$ using a transformation similar to
\begin{equation} 
  t' = t + \frac{2\dot{f}}{f_o}t^2,
\end{equation}
where $t$ is the time used when sampling the original data, \fdot\ is
the frequency derivative, and $f_o$ is the initial frequency of the
signal.  Additional details and variations on the theme can be found
in \citet{mk84, agk+90, jk91} and \citet{whn+91}.

By stretching the data using the appropriate transform and then FFTing
the corrected time series, we can recover the fully coherent response.
Such techniques have been used with significant success in searches of
relatively short time-series \citep[e.g.][]{clf+00}.  However, this
technique runs into significant difficulties when trying large numbers
of transformations using long time series where computation of the FFT
is non-trivial.  Both techniques that we will mention allow full
corrections to be made to a signal without requiring multiple FFTs of
the full original data set.

\subsubsection{Correlation Technique} \label{sec:correlation}

The correlation technique is the more powerful of the two methods, and
uses matched filtering in the Fourier domain to ``sweep-up'' signal
spread over a number of frequency bins into a single bin.  In
astrophysical applications, we usually have some sort of ``pure''
signal (like a harmonic from a millisecond pulsar), whose frequency
changes as a function of time due to some other process (such as
orbital motion or pulsar spin-down).  In the Fourier domain, these
processes cause the perfect $\sinc$-like response of a harmonic to be
spread over numerous local Fourier bins --- in effect, the $\sinc$
response is convolved with a finite impulse response (FIR) filter
(where finite in this case refers to a small portion, say $m$ bins, of
the frequency range analyzed rather than a short period of time).  If
we can predict the complex form (and phase) of that FIR filter, we can
recover the coherent response (i.e.\ the perfect $\sinc$-function) by
correlating the appropriate Fourier bins with a ``frequency-reversed''
and complex-conjugated template that matches the filter.

In mathematical terms, consider a signal with a normalized Fourier
response of $A_{k-r_o}$, where $k-r_o$ is simply the frequency offset
of bin $k$ from some reference frequency $r_o$, which goes to zero as
$\left|k-r_o\right|$ approaches some number of bins $m/2$.  For
Fourier interpolation as described in \S\ref{sec:interpolation}, this
response is equal to eqn.~\ref{eq:cosresp} without the $A_o$ factor
(i.e.\ normalized to an amplitude of one for a coherent response).
The complex-valued Fourier response of such a signal at frequency
$r_o$ can be calculated with the sum
\begin{equation}
  \label{eq:correlation}
  A_{r_o} \simeq \sum_{k=[r_o]-m/2}^{[r_o]+m/2} A_k\ A_{r_o-k}^*.
\end{equation}
If $r_o$ is initially unknown (i.e.\ we are searching for a signal
with the response shape as defined by the template but at an unknown
frequency) we simply compute this summation at a range of frequencies
$r$.

Calculating eqn.~\ref{eq:correlation} over a range of evenly spaced
frequencies is equivalent to correlating the raw FFT amplitudes with
the template and is therefore most efficiently computed using short
FFTs and the convolution theorem.  With FFTs of length $M$, such that
$m \ll M \ll N/2$, we can search a very-long FFT of length $N/2$ for
any signal whose $A_{k-r_o}$ we can compute, using overlap-and-save or
overlap-and-add techniques \citep[see e.g.][]{ptvf92}.  Such
calculations have advantages over standard time-domain ``stretching''
techniques in that that they are memory local and can be easily
parallelized --- important properties when dealing with very-long
time-series and modern distributed memory computer architectures.

Moving to our example of a signal with a constant frequency
derivative, a single harmonic of the signal has the form
\begin{mathletters}
  \begin{eqnarray}
  n(u) 
  & = & a\ \cos\left[2\pi\left(r_o u + \frac{\dot{r}}{2}u^2\right) + 
    \phi\right] \\
  & = & \frac{a}{2}\left[\me^{2\pi\mi\left(r_o u + 
        \frac{\dot{r}}{2}u^{2}\right)}
    \me^{\mi\phi}+\me^{-2\pi\mi\left(r_o u + \frac{\dot{r}}{2}u^{2}\right)}
    \me^{-\mi\phi}\right],
\end{eqnarray}
\end{mathletters}%
where we use the same notation as \S\ref{sec:dft}.  Neglecting the
second term as in \S\ref{sec:sinusoid} and Fourier transforming at
some ``center'' or average frequency $r'_c = r + \dot{r}/2$, we get
\begin{equation}
  A_{r'_c}=\frac{aN}{2}\me^{\mi\phi}\int^{1}_{0}
  \me^{\mi\pi\left(\dot{r}u^{2} + 2q_r u \right)}\,\mathrm{d}u,
\end{equation}
where $q_r=r_c-r'_c$, and the real ``center'' frequency of the signal
is $r_c=r_o+\dot{r}/2$.  This integral can be evaluated in closed form
\begin{equation}
  \int^{1}_{0}\me^{\mi\pi\left(\dot{r}u^{2}+2q_r u\right)}\,\mathrm{d}u = 
  \frac{1}{\sqrt{2\dot{r}}}\me^{-\mi\pi\frac{q_r^{2}}{\dot{r}}}
  \left\{S\left(Z_r\right)-S\left(Y_r\right)+
    \mi\left[C\left(Y_r\right)-C\left(Z_r\right)\right]\right\}
\end{equation}
where $Y_r=\sqrt{2/\dot{r}}q_r,$
$Z_r=\sqrt{2/\dot{r}}\left(q_r+\dot{r}\right),$ and $C(x)$ and $S(x)$
are the Fresnel integrals
\begin{equation}
  C(x)=\int_{0}^{x}\cos\left(\frac{\pi}{2}t^{2}\right)\,\mathrm{d}t,\ \
  S(x)=\int_{0}^{x}\sin\left(\frac{\pi}{2}t^{2}\right)\,\mathrm{d}t.
\end{equation} 
The Fourier transform response then becomes
\begin{equation}
  \label{eq:ffdotresp}
  A_{r'_c}=\frac{aN}{2\sqrt{2\dot{r}}} \ 
  \me^{\mi\left(\phi-\pi\frac{q_r^{2}}{\dot{r}}\right)} \
  \left\{S\left(Z_r\right)-S\left(Y_r\right)+
    \mi\left[C\left(Y_r\right)-C\left(Z_r\right)\right]\right\}.
\end{equation}

Using the correlation technique, the coherent response can be
recovered by convolving local Fourier amplitudes with the
``frequency-reversed'' and complex-conjugated template as defined by
eqn.~\ref{eq:ffdotresp}.  This response, at average Fourier frequency
$r_c$ and Fourier frequency derivative $\dot{r}$ can therefore be
written as
\begin{equation}
  \label{eq:ffdotcorr}
  A_{r_c,\dot{r}}=\sum_{k=\left[r\right]-m/2}^{k=\left[r\right]+m/2}A_{k}
  \ \frac{1}{\sqrt{2\dot{r}}}\left(\me^{\mi\pi\frac{q_k^2}{\dot{r}}}
    \left\{S\left(Z_k\right)-S\left(Y_k\right)-
      \mi\left[C\left(Y_k\right)-C\left(Z_k\right)\right]\right\}\right).
\end{equation}

Eqn.~\ref{eq:ffdotcorr} takes into account the fact that the signal
has been ``spread'' relatively evenly into the $\dot{r}$ closest
frequency bins to $r_{c}$, while an additional small amount of signal
has ``leaked'' into bins further away --- much like the non-zero wings
of the $\sinc$-like response to a constant frequency signal.  As a
rule of thumb, the correct Fourier amplitude will be well-approximated
if $m$ is chosen such that $\dot{r} < m \lesssim 2\dot{r}$.

Large scale searches of pulsations with constant frequency derivatives
have been conducted using the correlation technique.  A successful
example is a search for pulsars in globular cluster NGC~6544 using
data taken with the Parkes radio telescope \citep{rgh+01}.  The search
was conducted using an FFT of 13865600 points over Fourier $\dot{r}$
values from $-100$ to $100$ with a step-size of $\Delta\dot{r}=2$ and
included the calculation of amplitudes at half-bin frequency
intervals.  The search resulted in the detection of the 3.06\,ms
PSR~J1807-2459 in a low-mass binary with orbital period 1.7\,h, the
second shortest radio pulsar orbital period known.  A detailed view of
the pulsar's fundamental harmonic is shown in Fig.~\ref{fig:NGC6544}.
The plot was calculated using the correlation technique with spacings
of $\Delta r=0.0625$ and $\Delta\dot{r}=0.25$.  The generation of such
a piece of the \ffdot\ plane takes only a fraction of a second on a
rather modest workstation.

\subsubsection{Vector Bending} \label{sec:vector}

Vector bending is one of the simplest and most straightforward methods
to correct a Fourier response that has been smeared over several local
frequency bins.  As we described in \S\ref{sec:dft}, the DFT can be
thought of as the vector addition of $N$ complex numbers.  This
addition produces a straight line in the complex plane for a
coherently detected sinusoid.  For a sinusoid with a non-integral or
time-varying pulsation frequency, the standard DFT addition produces a
curved shape (see Fig.~\ref{fig:vectors}).  Since the amplitude of
the Fourier response is the distance between the origin and the
end-point of the vector addition, any curvature in the vector addition
implies non-optimal signal detection.

The precise shape of the response curve in the complex plane depends
on the mismatch of the signal's (possibly time-dependent) pulsation
frequency and the frequency used in forming the DFT addition (i.e.\ 
the closest FFT bin).  Regardless of the shape, though, for short
enough segments of the curve, the segments differ little from straight
lines.  We can therefore approximate the shape of the curve as a sum
of $G$ linear segments, each of which contains $N/G$ points from the
full-resolution vector addition.  In terms of the $r^{\rm th}$ DFT
amplitude, we can write this as
\begin{equation}
  \label{eq:fold}
  A_r = \sum_{g=0}^{G-1}B_{r,g} = \sum_{g=0}^{G-1} 
  \sum_{h=0}^{\frac{N}{G}-1}n_j\me^{-2\pi\mi jr/N},
\end{equation} 
where $j = gN/G+h$.  This is equivalent to calculating and then
summing $G$ independent DFTs (the $B_{r,g}$), each of which suffers
virtually no loss in sensitivity when the curvature over a segment is
small.

Using these vector addition segments or sub-vectors, we can correct
for the loss of sensitivity due to curvature by simply straightening
the vector addition.  If we can predict the true pulsation frequency
of a signal as a function of time, we can predict how much curvature
will accumulate in each sub-vector and then remove it by rotating the
segment appropriately.

For our example of a signal with constant frequency derivative, the
instantaneous phase (ignoring the intrinsic phase of the signal) is
equal to
\begin{mathletters}
  \begin{eqnarray}
    \Phi_{true}(u) 
    & = & 2\pi\int r(u)\,\mathrm{d}u \\
    & = & 2\pi\int\left(r_o + \dot{r}u \right)\,\mathrm{d}u \\
    & = & 2\pi r_o u + \pi\dot{r}u^2,
  \end{eqnarray}
\end{mathletters}%
where $r_o = f_o T$ is the initial pulsation frequency of the signal,
and $\dot{r} = \dot{f} T^2$ is the frequency derivative.  The process
of taking a DFT removes an instantaneous phase equivalent to
$\Phi_{DFT}(u) = 2\pi ru$ from the signal (see
eqn.~\ref{eq:ftintegral}).  So the instantaneous phase error is equal
to
\begin{mathletters}
  \begin{eqnarray}
    \label{eq:phierror}
    \Phi_{error}(u) 
    & = & \Phi_{true}(u) - \Phi_{DFT}(u) \\
    & = & 2\pi(r_o-r)u + \pi\dot{r}u^2.
  \end{eqnarray}
\end{mathletters}%
Therefore, to correct a particular signal using vector bending, we
first calculate the $B_{r,g}$ using eqn.~\ref{eq:fold} for a
particular Fourier frequency $r$ (such as the frequency of a known
pulsar).  Now we attempt to un-bend the full Fourier response by
summing the $B_{r,g}$ after correcting for the phase errors
$\Phi_{error}(u_g)$ as defined by eqn.~\ref{eq:phierror}.  The
corrected response is equal to
\begin{equation}
  A_{r,\dot{r}} = \sum_{g=0}^{G-1}B_{r,g}\me^{-\mi\Phi_{error}(u_g)}.
\end{equation}
A choice of $G\sim10^3$ will essentially eliminate the loss of
response for reasonable frequency offsets and frequency derivatives
(i.e. less than a few 10s of Fourier bins).

While impractical for large-scale searches due to the fact that the
$B_{r,g}$ must be recomputed every few $r$, vector bending offers
significant computational advantages in certain situations.  In
particular, X-ray observations often consist of short ($<1$ hour)
``on-source'' segments separated by hours, days, or even weeks of
``off-source'' time (see also \S\ref{sec:photon}).  An FFT of the
entire time series might be prohibitively expensive.  However, if we
can determine an ``initial guess'' frequency (e.g.\ by FFTing one
segment of the observation or from an ephemeris), we can quickly
calculate the $B_{k,g}$ at this frequency from the ``on-source''
intervals alone.  We can reconstruct the \ffdot\ plane around our
frequency of interest without ever creating the full ``filled-in'' or
padded time series, let alone calculating a potentially huge FFT.
Such techniques have allowed us to perform frequency analysis of
very-long stretches of data from ROSAT observations of PSR~B0540-59
\citep{efr98}.  Similarly, Fig.~\ref{fig:vectors} shows the use of the
method for a 2.4\,day observation of the Crab pulsar.

\section{Signal Property Estimation} \label{sec:sigprops}

Besides correcting for losses of sensitivity, Fourier interpolation
and the other response correcting techniques mentioned in
\S\ref{sec:respcorr} allow us to determine other useful properties of
a detected signal.  Using detailed amplitude and phase information
from a signal's Fourier harmonics, we can estimate properties such as
the statistical significance of the signal, the location and duration
of the signal in the time series (the ``centroid'' and ``purity''
respectively), the precise pulsation frequency, and estimates of the
measurement errors for Fourier power and phase.

\begin{figure}
  \plotone{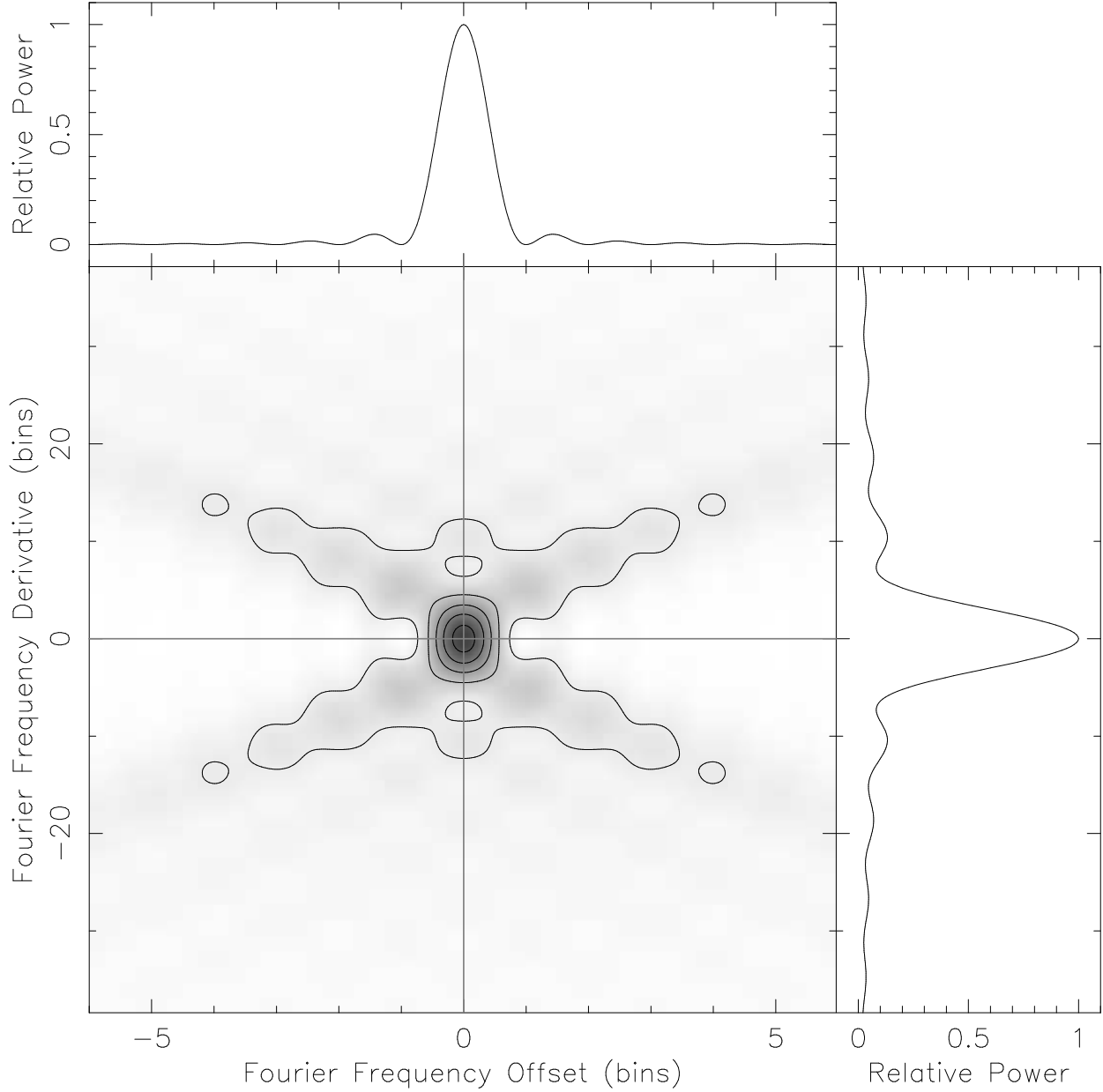}
  \caption{\footnotesize
    The theoretical response for a boxcar-windowed signal with a
    constant frequency derivative.  Contours plotted are 10\%, 30\%,
    50\%, 70\%, and 90\% of peak (i.e.\ fully corrected) power.  The
    top panel shows the familiar $\sinc$ response of the signal along
    the $\dot{f}=0$ line.  The right panel shows a similar cut along
    the $\Delta r=0$ line (i.e.\ the calculated average frequency is
    the true average signal frequency).  The relatively uniform spread
    of signal power over the local power spectrum is apparent for all
    values of frequency derivative.
    \label{fig:ffdot}}
\end{figure}

The first step when estimating signal properties in the Fourier domain
is to isolate the true peak of the Fourier response in power.  This is
easily accomplished by using the matched filtering techniques to
generate an oversampled grid of amplitudes near and around the signal
candidate (see Figs.~\ref{fig:ffdot} and \ref{fig:NGC6544}).  Simple
optimization algorithms such as the downhill simplex method can then
be used to refine the peak location \citep[e.g.][]{ptvf92}.  Once the
peak has been located, estimates of the first and second derivatives
of power and phase with respect to Fourier frequency, obtained using
Fourier interpolation, can be used to calculate various useful signal
properties \citep{mdk93}.

\placefigure{fig:ffdot}

\subsection{Power, Phase and Signal Amplitude}

When the peak of the Fourier response has been located as a function
of Fourier frequency and the other search parameters, the measured
power is defined as
\begin{equation}
  \label{eq:power}
  P_{meas} = \frac{\left|A_{r,\dots}\right|^2}{P_{norm}},
\end{equation}
where $P_{norm}$ is the expected noise power and is usually described
by one of $N\left< d_j^2 \right>$, $P_{local}$, or $n_{ph}$ as
discussed in \S\ref{sec:noise}.  \citet[see
\S\ref{sec:signoise}]{gro75d} showed that since the measured power is
a random variable due to the presence of noise, its variance is
$2P_{signal}+1$, where $P_{signal}$ is the power caused by the signal.
Since we do not \emph{a priori} know the true signal power, a good
estimate for the variance of the measured power is simply
$2P_{meas}-1$ since $\left<P_{meas}\right> = P_{signal}+1$.

Using $P_{meas}$, as well as the knowledge that a sinusoid of
amplitude $a$ in a noisy time series produces a power with an
expectation value of $\left<P_{meas}\right> =
a^2N^2/\left(4P_{norm}\right)+1$ (see~\S\ref{sec:sinusoid}), we can
estimate the signal amplitude as
\begin{equation}
  \label{eq:amp_est}
  \left<a\right> = \frac{2}{N}\sqrt{P_{norm}\left(P_{meas}-1\right)}.  
\end{equation}
For binned data containing a signal with Fourier frequency $r$, the
measured power should be multiplied by $1/\sinc^2(\pi r/N)$ to correct
for the loss in sensitivity due to binning (see~\S\ref{sec:binned}).
\citet{vvw+94} provide detailed instructions on how to estimate upper
limits on pulsation amplitudes as well as estimates of the overall
sensitivity of a search.

The statistical significance of a signal is also determined by
$P_{meas}$.  The probability that noise caused a particular power to
equal or exceed $P_{meas}$ is given by $\me^{-P_{meas}}$
(eqn.~\ref{powprob} with $P' = P_{meas}$).  But for a search over
$N_{IFS}$ independent Fourier powers, the probability that at least
one of the noise powers exceeds $P_{meas}$ is given by
\begin{equation}
  \label{eq:search_prob}
  Prob(P_{noise} \geq P_{meas}) = 
  1 - \left(1 - \me^{-P_{meas}}\right)^{N_{IFS}}. 
\end{equation}
\citet{vvw+94} show how to use this information to set detection
thresholds that minimize the number of spurious candidate signals and
give high confidence that signals with powers above the thresholds are
real.

Using the real and imaginary parts of the peak Fourier response, we
can also calculate the phase of the sinusoidal signal as 
\begin{equation}
  \label{eq:phase}
  \phi_{meas} = \arctan\left[\frac{{\rm Im}\left(A_{r,\dots}\right)}{{\rm
      Re}\left(A_{r,\dots}\right)}\right]
\end{equation}
radians.  Using similar arguments as for the measured power, the
variance of the measured phase is approximately $1 /(2P_{meas}-1)$
radians.

\subsection{Signal Location and Duration in Time} \label{sec:cen+pur}

Astronomical observations of pulsations effectively consist of a
\emph{window} of on-source time where pulsations are present, and the
rest of the time when they are not.  For most of this paper we have
assumed that a signal is present throughout the observation as
evidenced by the limits of integration for eqn.~\ref{eq:ftintegral},
which in time-normalized units go from $0 \le u \le 1$, or
equivalently, from $0 \le t \le T$.  In effect, pulsations such as
that defined in eqn.~\ref{eq:cosine} are multiplied by a square window
function defined as 1 during the observation and 0 at all other times.
This window function is simply a property of the DFT and is due to the
finite duration of our observation.

It is possible, though, and for various reasons often likely, that a
signal we are observing turns on and off or varies in intensity during
an observation.  The behavior of the signal itself effectively defines
a new window function, $W(u)$.  By measuring the moments of this
window function with respect to time, we can determine approximately
where in our data a signal is located and for how long.

The approximate location of a signal in a time-normalized time series
is described by its centroid, $\hat{C} = \left<u\right> =
\left<t\right>/T$ which is proportional to the first moment of the
window function with respect to time.  More specifically,
\begin{equation}
  \label{eq;centroid_def}
  \hat{C} = \frac{\int^\infty_{-\infty}u\,W(u)\,\mathrm{d}u}
  {\int^\infty_{-\infty}W(u)\,\mathrm{d}u}.
\end{equation}
\citet{mc82} wrote this in terms of the measured Fourier response as
\begin{equation}
  \label{eq:centroid}
  \hat{C} = -\frac{1}{2 \pi}\frac{\partial\phi(r_o)}{\partial r}
  \pm \frac{1}{\sqrt{24P(r_o)}},
\end{equation}
where $P(r_o)$ and $\phi(r_o)$ is the phase measured at the peak of the Fourier
response (i.e.\ $\phi_{meas}$) and $r$ is the Fourier frequency (see
Appendix~\ref{app:centroid} for a derivation).  Signals present
throughout an observation have $\hat{C} = 1/2$ while those present in
only the first or second halves of the observation have $\hat{C} =
1/4$ or $\hat{C} = 3/4$ respectively.

The second moment of the window function with respect to time is
related to the moment of inertia of a function and can therefore be
used to estimate the root-mean-squared (rms) dispersion of the
pulsations in time about the centroid.  Using this information,
\citet{mc82} defined a parameter called the ``purity'' (and symbolized
by $\alpha$) as
\begin{equation}
  \label{eq:purity}
  \alpha = \frac{1}{\pi}\sqrt{-\frac{3}{2P(r_o)}
    \frac{\partial^2 P(r_o)}{\partial r^2}}
  \pm \frac{1}{\alpha\sqrt{10P(r_o)}},
\end{equation}
where $P(r_o)$ is the measured power at the the peak of the Fourier
response (i.e.\ $P_{meas}$, see Appendix~\ref{app:purity} for a
derivation).  The scaling in eqn.~\ref{eq:purity} is chosen such that
the rms dispersion of the signal about the centroid for a window
function $W(u)$, is equivalent to that of a rectangular window
function of duration $\alpha$ (in units of the time series length)
centered on the centroid.  A signal present throughout the data would
have $\alpha = 1$, while one present in only half the data (in a
continuous section) would have $\alpha = 1/2$.  Signals present only
at the start and end of an observation but absent in the middle have
$\alpha > 1$.  Purity can also help to identify sidelobes caused by a
periodic modulation of a signal as these Fourier amplitudes have
$\alpha = \sqrt{3}$.

Since the location and duration of a signal in an time series affects
the Fourier response, it is important to understand how
eqn.~\ref{eq:cosresp} changes if a signal is present during only part
of an observation.  In Appendix~\ref{app:response}, we show that when
close to the peak of a signal's Fourier response,
\begin{equation}
  \label{eq:cpresp}
  A_r \simeq A_o\,\me^{-2\pi\mi\hat{C}(r-r_o)}\,
  \sinc\left[\pi\alpha(r-r_o)\right]
\end{equation}
where $A_o = Na/2\;\me^{\mi \phi_o}$, $\phi_o$ is the intrinsic phase
of the signal, and $r_o$ is the true signal frequency (in FFT bins).
This equation demonstrates that for centroids different from $1/2$,
the phase shift between consecutive FFT bins differs from the $\pi$
radians shown in eqn.~\ref{eq:cosresp}.  Similarly, for purity values
different from 1, neighboring FFT bins show more or less correlation
with each other (i.e.\ the central peak of the \sinc\ function changes
its width).

\subsection{The Pulsation Frequency and Frequency Derivative}

The true pulsation frequency of the signal is located at the point
where $\partial P / \partial r = 0$.  Furthermore, given the response
in eqn.~\ref{eq:cpresp}, we can show (Appendix~\ref{app:freqsigma})
that the uncertainty in this measurement (in Fourier bins) is given by
\begin{equation}
  \label{eq:freqsigma}
  \sigma_r = \frac{3}{\pi\alpha\sqrt{6 P_{meas}}}.
\end{equation}
This uncertainty is considerably smaller than the often-quoted
``frequency error'' for the FFT of one bin width, which is simply the
frequency resolution returned by the FFT algorithm.

If the correlation method is used to isolate a peak in the \ffdot\ 
plane as shown in Figs~\ref{fig:ffdot} and \ref{fig:NGC6544}, we can
calculate the uncertainty in the measured $\dot{r}$ value by using
similar arguments and methods as for the frequency uncertainty (see
Appendix~\ref{app:fdotsigma} for a derivation).  The uncertainty in
the $\dot{r}$ (in Fourier bins) is approximately
\begin{equation}
  \label{eq:fdoterr}
  \sigma_{\dot{r}} = \frac{3 \sqrt{10}}{\pi\alpha^2\sqrt{P_{meas}}}.
\end{equation}

\section{Conclusions} \label{sec:conclusions}

In this paper we have described techniques that allow sophisticated
and fully coherent Fourier analysis of very long time series.  Most of
these techniques use the wealth of information provided by the Fourier
phases --- information discarded during ``standard'' analyses based on
raw power spectra.

Significant gains in sensitivity and efficiency are possible when
using Fourier phase information during the search for periodic signals
(using the Fourier domain matched filtering techniques described in
\S\ref{sec:respcorr}) and when characterizing signals that are known
to be present in the data (using the parameters described in
\S\ref{sec:sigprops}).  The methods of Fourier domain matched
filtering allow efficient, memory local, and inherently parallel
analysis of extremely long time series with only modest computational
resources.  Billion point out-of-core FFTs followed by fully coherent
matched filtering pulsation searches are possible on ``standard''
workstations.  More traditional time domain based techniques (such as
acceleration searches performed by stretching or compressing the time
series followed by large in-core FFTs) on similarly sized time series
require specialized high-performance computing resources, assuming
they can be performed at all.

As astronomical instruments become more sophisticated and specialized,
time series of ever increasing duration and time resolution will
appear.  The Fourier domain techniques described in this paper should
prove to be essential tools in their analysis.

\emph{Acknowledgments}\ \ We would like to thank G.~Fazio for
supporting our research and encouraging us in this work.  Additional
thanks go to J.~Grindlay, B.~Schmidt, F.~Seward, R.~Narayan, V.~Kaspi,
and L.~Greenhill for their encouragement and comments.  Many of the
computations for this paper were performed on equipment purchased with
NSF grant PHY 9507695.  S.M.R. acknowledges the support of a Tomlinson
Fellowship awarded by McGill University.  S.S.E. is supported in part
by an NSF CAREER Grant.

\appendix

\section{Derivation of Fourier Interpolation} \label{app:fourinterp}

Following the derivation found in \citet{mdk93}, we begin with the
definition of the $k^{\rm th}$ DFT element
\begin{equation} 
A_k = \sum_{j=0}^{N-1} n_j \ \me^{-2\pi\mi jk/N}, 
\end{equation} 
which we then rewrite by substituting the inverse DFT for the $n_j$
\begin{mathletters}
  \begin{eqnarray}
    A_k 
    & = & \sum_{j=0}^{N-1} \left(\frac{1}{N}\sum_{l=0}^{N-1} 
      A_l \ \me^{2\pi\mi jl/N}\right) \me^{-2\pi\mi jk/N} \\
    \label{appb1}
    & = & \frac{1}{N} \sum_{l=0}^{N-1} A_l 
    \sum_{j=0}^{N-1} \me^{-2\pi\mi j(k-l)/N}.
  \end{eqnarray}
\end{mathletters}%
The last summation can be computed exactly using the identity
\begin{equation}
  \label{eq:sumidentity}
  \sum_{j=0}^{N-1} \me^{\mi\alpha j} = \me^{\frac{\mi\alpha}{2}(N-1)}
  \frac{\sin\left(\frac{N\alpha}{2}\right)}
  {\sin\left(\frac{\alpha}{2}\right)},
\end{equation}
such that when $N\gg1$ we have
\begin{mathletters}
  \begin{eqnarray}
    \sum_{j=0}^{N-1} \ \me^{-2\pi\mi j(k-l)/N} 
    & = & \me^{-\mi\pi(k-l)(1-\frac{1}{N})}
    \frac{\sin\left[\pi(k-l)\right]}{\sin\left[\pi(k-l)/N\right]} \\
    & \simeq & \me^{-\mi\pi(k-l)}\frac{\sin\left[\pi(k-l)\right]}{\pi(k-l)/N} \\
    \label{appb2}
    & \simeq & N\ \me^{-\mi\pi(k-l)} \sinc\left[\pi(k-l)\right].
  \end{eqnarray}
\end{mathletters}%
Substituting this expression into eqn.~\ref{appb1} and changing the
integer frequency $k$ into a continuous real-valued frequency $r$, we
arrive at eqn.~\ref{eq:conv1}
\begin{equation} 
  A_r = \sum_{l=0}^{N-1}A_l\ \me^{-\mi\pi(r-l)} 
  \sinc\left[\pi(r-l)\right].
\end{equation}

\section{Derivation of Frequency Uncertainty} \label{app:freqsigma}

For a given Fourier frequency offset $\Delta_r = r - r_o$, where $r_o$
is the Fourier frequency of the signal, the magnitude of the Fourier
response and the power go as (see Appendix~\ref{app:response})
\begin{mathletters}
  \begin{eqnarray}
    \left|A(r)\right| 
    & = & \left|A_o\right|\sinc\left(\pi\alpha\Delta_r\right) \\
    P(r)
    & = & P_{meas} \ \sinc^2\left(\pi\alpha\Delta_r\right)
  \end{eqnarray}
\end{mathletters}%
where $\alpha$ is the signal purity.  We can expand the $\sinc$
function in order to approximate the expression for the power near the
peak of the response as
\begin{equation}
  \label{eq:powexpand1}
  P(r) = P_{meas} \left[\frac{\sin(\pi\alpha\Delta_r)}
    {\pi\alpha\Delta_r}\right]^2 
  \simeq P_{meas} \left[1 - 
    \frac{\left(\pi\alpha\Delta_r\right)^2}{3}\right].
\end{equation}
Taking the derivative of power with respect to $r$ and solving for
$\Delta_r$ we obtain
\begin{equation}
  \Delta_r = -\frac{3}{2\pi^2\alpha^2P_{meas}}
  \frac{\partial P}{\partial r}.
\end{equation}
As expected, when the Fourier frequency equals the true frequency of
the pulsations (i.e.\ $\Delta_r = 0$), the Fourier response peaks and
$\partial P(r)/\partial r = 0$.

In order to estimate the uncertainty in $\Delta_r$, we apply standard
propagation of errors to arrive at
\begin{equation}
  \label{eq:sigr}
  \sigma_r = \frac{3}{2\pi^2\alpha^2P_{meas}} \sigma_{P'(r)},
\end{equation}
where we have replaced $\partial P(r)/\partial r$ with $P'(r)$
to simplify the notation.  The derivative of the power at the true
frequency can be approximated using finite differences as
\begin{equation}
  \label{eq:dpow_diff}
  P'(r) \simeq \frac
  {P\left(r_o + \Delta_r\right) - P\left(r_o - \Delta_r\right)}
  {2\Delta_r} = \frac{P^+ - P^-}{2\Delta_r},
\end{equation}
where we have simply renamed $P\left(r_o + \Delta_r\right)$ and
$P\left(r_o - \Delta_r\right)$.  The uncertainty in $P'(r)$ can also
be approximated using finite differences and error propagation.  Since
$P^+$ and $P^-$ are highly correlated when $\Delta_r \ll 1$, their
uncertainties are also correlated giving
\begin{equation}
    \label{eq:sigpp}
    \sigma_{P'(r)} 
    \simeq \frac{1}{2 \Delta_r} \left(\sigma_{P^+} + \sigma_{P^-}\right) 
    \simeq \frac{\sigma_{P^+}}{\Delta_r}.
\end{equation}

Now, we turn to the question of the uncertainty in $P^+$, closely
following \citet{mid76}.  The amplitude of the Fourier response at the
true signal frequency can be written as
\begin{equation} 
  \label{eq:freq_pexpand}
  \sqrt{P_{meas}} = \sum_{j=0}^{N-1} y_j \cos(\phi_j)
\end{equation}
where $y_j = n_j/\sqrt{P_{norm}}$ are the points in our time series as
defined in eqn.~\ref{ftdef}, but scaled using the appropriate
$P_{norm}$ such that the measured power, $P_{meas}$, is properly
normalized (see~\S\ref{sec:noise}).  Similarly, the $\phi_j$ represent
the pulsation phases at times $u = j/N$, but rotated by the measured
Fourier phase, $\phi_{meas}$, such that the result of the vector
addition lies along the real axis in the complex plane (i.e.\ the
final complex phase is zero).  In effect, this transform isolates
components of the data that are parallel to the final Fourier
response.

At a small frequency offset $\Delta_r$ from the true frequency, we can
expand the power in a similar fashion as
\begin{equation}
  \label{eq:powexpand}
  \sqrt{P^+} = \sum_{j=0}^{N-1} y_j 
  \cos\left(\phi_j + \delta_{\phi_j}\right),
\end{equation}
where $\delta_{\phi_j}$ are the ``phase errors'' introduced by the
frequency offset.  The phase errors add curvature to the vector
addition and are defined as
\begin{equation}
  \label{eq:deltadef}
    \delta_{\phi_j} = 
    2\pi\alpha\Delta_r\left(\frac{j}{N}-\frac{1}{2}\right) = 
    2\pi\alpha\Delta_r\left(u-\frac{1}{2}\right),
\end{equation}
where $u$ is the normalized time, $u=t/T=j/N$. The $1/2$ term in
eqn.~\ref{eq:deltadef} removes the accumulated phase error over the
course of the observation (i.e.\ $\int_0^1 2\pi\alpha\Delta_r
u\,\mathrm{d}u = \pi\alpha\Delta_r$) and makes the vector summation of
eqn.~\ref{eq:freq_pexpand} finish on the real axis.  Expanding the
cosine in eqn.~\ref{eq:powexpand} gives
\begin{equation}
  \label{eq:powexpand2}
  \sqrt{P^+} = 
  \sum_{j=0}^{N-1} y_j \cos(\phi_j)\cos\left(\delta_{\phi_j}\right) - 
  \sum_{j=0}^{N-1} y_j \sin(\phi_j)\sin\left(\delta_{\phi_j}\right).
\end{equation}

Considering the uncertainties in the separate terms of
eqn.~\ref{eq:powexpand2}, since the cosine term is an even function of
$\delta_{\phi_j}$, it is symmetric about $r_o$, and therefore does not
contribute to the uncertainty in the $P'(r)$ measurement as defined by
eqn.~\ref{eq:dpow_diff}.  For the sine term, given that the
$\cos(\phi_j)$ ``derotates'' the signal onto the real axis by
definition, we see that
\begin{equation}
  \sum_{j=0}^{N-1} y_j \sin(\phi_j) = 0.
\end{equation}
Furthermore, since the $\phi_j$ and $\delta_{\phi_j}$ are
uncorrelated, the average value of this term will be zero,
\begin{equation}
    \left<\sum_{j=0}^{N-1} y_j \sin(\phi_j)
      \sin\left(\delta_{\phi_j}\right)\right> = 0,
\end{equation}
and has no systematic effect on $P^+$.  However, we can
calculate the fluctuations introduced by this term
\begin{equation}
  \label{eq:sine_fluct}
  \left<\left(\sum_{j=0}^{N-1} y_j \sin\left(\phi_j\right)
      \sin\left(\delta_{\phi_j}\right)\right)^{2}\right> = 
  \left<\left(\sum_{j=0}^{N-1} y_j \sin\left(\phi_j\right)\right)^{2}\right> 
  \left<\sin^{2}\left(\delta_{\phi_j}\right)\right>.
\end{equation}
where the cross-terms average to zero since $\phi_j$ and
$\delta_{\phi_j}$ are uncorrelated.  Due to the normalization of the
$y_j$, the sum component averages to $1/2$.  The $\delta_{\phi_j}$
component has an average of
\begin{mathletters}
  \begin{eqnarray}
    \left<\sin^2\left(\delta_{\phi_j}\right)\right>
    & = & \int_0^1 \sin^2\left(\delta_{\phi_j}\right)\,\mathrm{d}u\\
    & \simeq & \int_0^1 \delta_{\phi_j}^2\,\mathrm{d}u\\
    & \simeq & \int_0^1 (2\pi\alpha\Delta_r)^2
    \left(u^2-u+\frac{1}{4}\right)\,\mathrm{d}u\\
    & \simeq & \frac{(\pi\alpha\Delta_r)^2}{3}.
  \end{eqnarray}
\end{mathletters}%
Therefore eqn.~\ref{eq:sine_fluct} is equal to $(\pi\alpha\Delta_r)^2/6$,
and the variance of $\sqrt{P^+}$ will be
\begin{equation}
  \label{eq:amp_variance}
  \sigma_{\sqrt{P^+}}^2 = \sqrt{P^+}\frac{(\pi\alpha\Delta_r)^2}{6}.
\end{equation}
To get the standard deviation we take the square root of this
expression.  Propagating errors to get the uncertainty in $P^+$ adds a
factor of $2\sqrt{P^+}$ to give
\begin{equation}
  \sigma_{P(r_o+\Delta_r)} = 2\pi\alpha\Delta_r\sqrt{\frac{P_{meas}}{6}},
\end{equation}
where we have used the approximation $P^+ \simeq P_{meas}$.  Finally,
substituting into equations~\ref{eq:sigpp} and~\ref{eq:sigr}, we have
\begin{equation}
  \label{eq:rerror}
  \sigma_r = \frac{3}{\pi\alpha\sqrt{6 P_{meas}}}.
\end{equation}

A much simpler derivation of eqn.~\ref{eq:rerror} is possible if we
realize that properly normalized powers times two are distributed
according to a $\chi^2$ distribution with 2 degrees of freedom
(see~\S\ref{sec:noise}).  For such a distribution, a $1\sigma$ error
corresponds to a change in the measured $\chi^2$ of $1/2$.  In the
case of powers, the $\pm1\sigma$ errors can therefore be found by
starting with the expansion of the power around $P(r_o)=P_{meas}$ as
given in eqn.~\ref{eq:powexpand1},
\begin{equation}
  P_{meas} \left[1 - \frac{\left(\pi\alpha\Delta_r\right)^2}{3}\right] = 
  P_{meas} - \frac{1}{2},
\end{equation}
and then solving for the $\Delta_r$ that corresponds to $P_{meas} -
1/2$.  This gives
\begin{equation}
  \Delta_{r} = \sigma_r = \frac{3}{\pi\alpha\sqrt{6 P_{meas}}}.
\end{equation}

\section{Derivation of $\dot f$ Uncertainty} \label{app:fdotsigma}

If a peak in the \ffdot\ plane has been isolated using techniques
similar to those shown in \S\ref{sec:correlation}, we can calculate
the error in the measurement of the true Fourier frequency derivative
$\dot{r_o}=\dot{f_o}T^2$ in a manner similar to that for the frequency
uncertainty as described in Appendix~\ref{app:freqsigma}.  Signals
with non-zero frequency derivatives have Fourier peaks which are
located off the $\dot{r} = 0$ line in the \ffdot\ plane, but the
\emph{shapes} of those peaks are independent of $\dot{r}$ and in fact
depend only on the window function of the signal (see
Appendix~\ref{app:response}).  The shape of the response in power as a
function of $\Delta_{\dot{r}} = \dot{r} - \dot{r_o}$ at the
``correct'' Fourier frequency $r = r_o$ is described by
eqn.~\ref{eq:ffdotresp}, and can be written as
\begin{equation}
  A(r_o, \dot{r}) = A_o\ \me^{-\mi\pi\frac{\Delta_{\dot{r}}}{4}}
  \sqrt{\frac{2}{\alpha^2\Delta_{\dot{r}}}}\left[
    S\left(\sqrt{\frac{\alpha^2\Delta_{\dot{r}}}{2}}\right) - \mi
    C\left(\sqrt{\frac{\alpha^2\Delta_{\dot{r}}}{2}}\right)\right],
\end{equation}
when $q_r$ is defined as
\begin{equation}
  q_r = r_c - r'_c = 
  \left(r_o + \frac{\dot{r_o}}{2}\right) - 
  \left(r_o + \frac{\dot{r_o} + \alpha^2\Delta_{\dot{r}}}{2}\right) = 
  -\frac{\alpha\Delta_{\dot{r}}}{2}.
\end{equation}
This definition of $q_r$ keeps the magnitude of the Fourier response
symmetric about $q_r$ no matter what the value of $\Delta_{\dot{r}}$.
The power as a function of $\Delta_{\dot{r}}$ is therefore
\begin{equation}
  P(r_o, \dot{r}) = P_o\frac{2}{\alpha^2\Delta_{\dot{r}}}\left\{
    \left[S\left(\sqrt{\frac{\alpha^2\Delta_{\dot{r}}}{2}}\right)\right]^2 +
    \left[C\left(\sqrt{\frac{\alpha^2\Delta_{\dot{r}}}{2}}\right)\right]^2\right\}.
\end{equation}

If we expand the Fresnel integrals about $0$ as
\begin{equation} 
  C(x) \simeq x - \frac{\pi^{2}}{40}x^{5}\dots, \ \hbox{and} \ 
  S(x) \simeq \frac{\pi}{6}x^{3}\dots,
\end{equation}
and then substitute, the power becomes
\begin{equation}
  P(r_o, \dot{r}) \simeq P_o 
  \left[1 - \frac{(\pi\alpha^2\Delta_{\dot{r}})^2}{180}\right].
\end{equation}
Taking the derivative of $P(r_o, \dot{r})$ with respect to
$\dot{r}$ and solving for $\Delta_{\dot{r}}$ gives 
\begin{equation}
  \Delta_{\dot{r}} = -\frac{90}{\pi^2\alpha^4P_o}
  \frac{\partial P(r_o, \dot{r})}{\partial \dot{r}}.
\end{equation}

From propagation of errors, 
\begin{equation}
  \label{eq:fdot_sigrdot}
  \sigma_{\dot{r}} = -\frac{90}{\pi^2\alpha^4P_o}\sigma_{P'(r_o, \dot{r})},
\end{equation}
where $P'(r_o, \dot{r}) = \partial P(r_o, \dot{r})/ \partial \dot{r}$.
Similarly, after using a finite difference estimate of the power
derivative following Appendix~\ref{app:freqsigma}, we find
\begin{equation}
  \label{eq:fdot_sigpderiv}
  \sigma_{P'(r_o, \dot{r})} = 
  \frac{\sigma_{P(r_o, \Delta_{\dot{r}})}}{\Delta_{\dot{r}}},
\end{equation}
where $P(r_o, \Delta_{\dot{r}})$ represents the power as measured at
$\dot{r} = \dot{r_o} + \Delta_{\dot{r}}$.

Closely following Appendix~\ref{app:freqsigma}, we represent
$\sqrt{P(r_o, \Delta_{\dot{r}})}$ as a sum of the parallel components
of the properly normalized time series points, $y_j$, as
\begin{equation}
  \label{eq:fdot_pexpand}
  \sqrt{P(r_o, \Delta_{\dot{r}})} = \sum_{j=0}^{N-1} y_j 
  \cos\left(\phi_j + \delta_{\phi_j}\right).
\end{equation}
The $\delta_{\phi_j}$ are the ``phase errors'' introduced when
$\Delta_{\dot{r}} \ne 0$, and are defined as
\begin{equation}
  \delta_{\phi_j} = \pi\alpha^2\Delta_{\dot{r}}\left(u^2-u+\frac{1}{6}\right),
\end{equation}
where the $u$ term comes from keeping the response symmetric about
$r_o$ (i.e.\ $q_r = -\alpha^2\Delta_{\dot{r}}/2$) and the $1/6$
removes the accumulated phase error over the course of the observation
(i.e.\ $\int_0^1 \pi\alpha^2\Delta_{\dot{r}}\left(u^2 -
  u\right)\,\mathrm{d}u = -\pi\alpha^2\Delta_{\dot{r}}/6$) and makes
the vector summation of eqn.~\ref{eq:fdot_pexpand} finish on the real
axis.

Expanding the cosine term of eqn.~\ref{eq:fdot_pexpand} gives
\begin{equation}
  \sqrt{P(r_o, \Delta_{\dot{r}})} = 
  \sum_{j=0}^{N-1} y_j 
  \cos\left(\phi_j\right)\cos\left(\delta_{\phi_j}\right) -
  \sum_{j=0}^{N-1} y_j 
  \sin\left(\phi_j\right)\sin\left(\delta_{\phi_j}\right).
\end{equation}
The first term shortens both $P(r_o, \Delta_{\dot{r}})$ and $P(r_o, -
\Delta_{\dot{r}})$ by the same amount and does not affect the
derivative of power.  For the sine term, since $\phi_j$ and
$\delta_{\phi_j}$ are uncorrelated, the average value is zero (see
Appendix~\ref{app:freqsigma}), but its fluctuations are important.  To
calculate the fluctuations, we square the terms and get
\begin{equation}
  \left<\left(\sum_{j=0}^{N-1} y_j \sin\left(\phi_j\right)
      \sin\left(\delta_{\phi_j}\right)\right)^{2}\right> = 
  \left<\left(\sum_{j=0}^{N-1} y_j \sin\left(\phi_j\right)\right)^{2}\right> 
  \left<\sin^{2}\left(\delta_{\phi_j}\right)\right>.
\end{equation}
Due to the normalization of the $y_j$, $\left<\left(\sum_{j=0}^{N-1}
    y_j \sin\left(\phi_j\right)\right)^{2}\right>$ averages to $1/2$
as before, and we can directly calculate
$\left<\sin^{2}\left(\delta_{\phi_j}\right)\right>$ as
\begin{mathletters}
  \begin{eqnarray}
    \left<\sin^{2}\left(\delta_{\phi_j}\right)\right> 
    & = & \int_{0}^{1}\sin^{2}\left(\delta_{\phi_j}\right)\,\mathrm{d}u \\
    & \simeq & \int_{0}^{1}\delta_{\phi_j}^{2}\,\mathrm{d}u \\
    & = & \pi^2\alpha^4\Delta_{\dot{r}}^2 
    \int_{0}^{1}\left(u^{2}-u+\frac{1}{6}\right)^{2}\,\mathrm{d}u \\
    & = & \frac{\pi^2\alpha^4\Delta_{\dot{r}}^2}{180}.
  \end{eqnarray}
\end{mathletters}%
The fluctuations from the sine term are therefore
$(\pi\alpha^2\Delta_{\dot{r}})^2/360$, and since squaring
eqn.~\ref{eq:fdot_pexpand} doubles the errors, the standard deviation
of $P(r_o, \Delta_{\dot{r}})$ is
\begin{equation}
  \sigma_{P(r_o, \Delta_{\dot{r}})} =
  \sqrt{P(r_o, \Delta_{\dot{r}})} \ \frac{2\pi\alpha^2\Delta_{\dot{r}}}{\sqrt{360}} 
  \simeq \sqrt{P_{meas}} \ \frac{\pi\alpha^2\Delta_{\dot{r}}}{\sqrt{90}}.
\end{equation}
Substituting into equations~\ref{eq:fdot_sigpderiv} and
then~\ref{eq:fdot_sigrdot} as in Appendix~\ref{app:freqsigma} gives us
the uncertainty in the frequency derivative
\begin{equation}
  \sigma_{\dot{r}} = \frac{3 \sqrt{10}}{\pi\alpha^2\sqrt{P_{meas}}}.
\end{equation}

\section{Derivation of Centroid} \label{app:centroid}

The centroid is a measure of the approximate location of a signal in a
time series as estimated by the first moment of the signal with
respect to time (see~\S\ref{sec:cen+pur}).  We can think of a
sinusoidal signal in our data as being always present but modulated in
intensity by some window function $W(u)$, where $u = t/T$ is the
normalized time and $T$ is the length of the observation.  A
``normal'' observation of a pulsar of constant intensity would
therefore have $W(u)=1$ when $0<u<1$ and $W(u)=0$ at all other times
(i.e.\ a square window).  Our signal is therefore described by
\begin{equation}
  \label{eq:cent_sigdef}
  s(u) = a\cos\left(2\pi r_o u + \phi_o\right)W(u),
\end{equation}
where $a$ is the amplitude, $r_o = f_oT$ is the Fourier frequency, and
$\phi_o$ is the phase of the sinusoid at time $u=0$.

Since the centroid of a function is proportional to the first moment
of the function with respect to time, we can easily calculate the
centroid using the Moment Theorem of Fourier transforms.
\citet{bra99} does this and defines the centroid as
\begin{equation}
  \label{eq:true_centroid}
  \left<u\right> = -\frac{A'(0)}{2\pi\mi A(0)},
\end{equation}
where $A(0)$ and $A'(0)$ are the Fourier transform and its first
derivative with respect to $r$ measured at $r=0$.

Eqn.~\ref{eq:true_centroid} is not directly applicable for our
sinusoidal signal since the information about the window function in
eqn.~\ref{eq:cent_sigdef} has been shifted to the frequency of the
sinusoid in accordance with the Modulation Theorem of Fourier
transforms.  Accordingly, we can apply the Modulation Theorem to
eqn.~\ref{eq:true_centroid} which gives us
\begin{equation}
  \label{eq:mod_centroid}
  \left<u\right>_{s(u)} = -\frac{A'(r_o)}{2\pi\mi A(r_o)}.
\end{equation}

Finally, we can write $A(r)$ in phasor form as $A(r) =
a(r)\me^{\mi\phi(r)}$ where $a(r)$ and $\phi(r)$ represent the Fourier
amplitude and phase as functions of the Fourier frequency $r$.  The
derivative with respect to Fourier frequency can be written
\begin{equation}
  A'(r) = \frac{\partial a(r)}{\partial r} \me^{\mi \phi(r)} +
  a(r)\mi\frac{\partial \phi(r)}{\partial r} \me^{\mi \phi(r)}.
\end{equation}
At the frequency of our signal, the amplitude is $A(r_o) =
a(r_o)\me^{\mi\phi(r_o)}$ and the Fourier response is at its peak,
making $\partial a(r_o)/\partial r = 0$.  Therefore
\begin{equation}
  \label{eq:cent_aderiv}
  A'(r_o) = A(r_o)\mi\frac{\partial\phi(r_o)}{\partial r}.
\end{equation}
Substituting eqn.~\ref{eq:cent_aderiv} into eqn.~\ref{eq:mod_centroid}
we arrive at
\begin{equation}
  \hat{C} = 
  \left<u\right>_{s(u)} = 
  -\frac{1}{2\pi}\frac{\partial\phi(r_o)}{\partial r},
\end{equation}
which is equivalent to eqn.~\ref{eq:centroid}.

We can also estimate the uncertainty on the measured value of the
centroid.  Following Appendix~\ref{app:freqsigma} from
eqns.~\ref{eq:powexpand}$-$\ref{eq:amp_variance}, we note that the
same noise fluctuations introduced when offsetting from the true
pulsation frequency by a small amount $\Delta_r$ that effect the
Fourier amplitude will also effect the Fourier phases.  When
$\Delta_r$ is small and the powers are properly normalized, the
amplitude fluctuations of eqn.~\ref{eq:sine_fluct} with variance
$\pi^2\Delta_r^2/6$ correspond to an uncertainty in the phase
measurement of $\sigma_{\phi(r_o+\Delta_r)} \simeq
\pi\Delta_r/\sqrt{6P(r_o)}$\,radians and therefore
\begin{equation}
  \sigma_{\hat{C}} = -\frac{1}{2\pi}\frac{\sigma_{\phi(r_o+\Delta_r)}}{\Delta_r}
  \simeq \frac{1}{\sqrt{24P(r_o)}}.
\end{equation}

\section{Derivation of Purity} \label{app:purity}

The ``purity'' of a signal (see~\S\ref{sec:cen+pur}) is a measure of
the rms dispersion of the pulsations in time with respect to the
centroid and is directly proportional to the variance in time of the
window function of the sinusoid from eqn.~\ref{eq:cent_sigdef}.  The
time variance is defined as
\begin{equation}
  \label{eq:variance_def}
  \left<\left(u-\left<u\right>\right)^2\right> = 
  \frac{\int^\infty_{-\infty}\left(u-\left<u\right>\right)^2\,
    W(u)\,\mathrm{d}u}
  {\int^\infty_{-\infty}W(u)\,\mathrm{d}u},
\end{equation}
but can be written as 
\begin{mathletters}
  \begin{eqnarray}
    \label{eq:true_variance}
    \left<\left(u-\left<u\right>\right)^2\right> 
    & = & \left<u^2\right> - \left<u\right>^2 \\
    & = & -\frac{A''(0)}{4\pi^2 A(0)} + 
    \frac{\left[A'(0)\right]^2}{4\pi^2\left[A(0)\right]^2},
  \end{eqnarray}
\end{mathletters}%
using the Moment Theorem for Fourier transforms \citep[see
e.g.][]{bra99}.  Since our signal is sinusoidal (see
Appendix~\ref{app:centroid}), application of the Modulation Theorem
gives
\begin{mathletters}
  \begin{eqnarray}
    \left<\left(u-\left<u\right>\right)^2\right>_{s(u)}
    & = & -\frac{A''(r_o)}{4\pi^2 A(r_o)} + 
    \label{eq:mod_variance}
    \frac{\left[A'(r_o)\right]^2}{4\pi^2\left[A(r_o)\right]^2} \\
    & = & -\frac{A''(r_o)}{4\pi^2 A(r_o)} - \hat{C}^2.
  \end{eqnarray}
\end{mathletters}%

Using $A(r) = a(r)\me^{\mi \phi(r)}$ and remembering that
$\partial a(r_o)/\partial r = 0$, the second derivative of the
Fourier amplitude is
\begin{equation}
  \label{eq:aderiv2}
  A''(r_o) = \mi A(r_o)\frac{\partial^2\phi(r_o)}{\partial r^2} - 
  A(r_o)\left(\frac{\partial\phi(r_o)}{\partial r}\right)^2 +
  \frac{A(r_o)}{a(r_o)}\frac{\partial^2 a(r_o)}{\partial r^2}.
\end{equation}
From Appendix~\ref{app:centroid}, we know that
$\partial\phi(r_o)/\partial r = -2\pi\hat{C}$ which makes the first
term of eqn.~\ref{eq:aderiv2} equal to zero since
$\partial^2\phi(r_o)/\partial r^2 = 0$ and the second term equal to
$-4\pi^2A(r_o)\hat{C}^2$.  The second derivative of power at the peak
response can be written as
\begin{equation}
  \frac{\partial^2 P(r_o)}{\partial r^2} = 
  \frac{\partial^2}{\partial r^2}\left(A^*(r_o)A(r_o)\right) = 
  2a(r_o)\frac{\partial^2 a(r_o)}{\partial r^2},  
\end{equation}
making the third term equal to
$\frac{A(r_o)}{2\left(a(r_o)\right)^2}\frac{\partial^2
  P(r_o)}{\partial r^2}$.  Substituting into
eqn.~\ref{eq:mod_variance} and simplifying gives
\begin{equation}
  \label{eq:good_variance}
  \left<\left(u-\left<u\right>\right)^2\right>_{s(u)} =
  -\frac{1}{8\pi^2P(r_o)}\frac{\partial^2 P(r_o)}{\partial r^2}.
\end{equation}

If we normalize the variance using the value obtained for a signal
present throughout the observation (i.e.\ a square window where $W(u)
= 1$ from $0 \le u \le 1$ and zero elsewhere --- which we will call a
\emph{pure} signal) where
\begin{equation}
  \left<\left(u-\left<u\right>\right)^2\right>_{pure} = 
  \left<u^2\right> - \left<u\right>^2 = 
  \frac{1}{3} - \left(\frac{1}{2}\right)^2 = \frac{1}{12},
\end{equation}
and then taking the square-root, we are left with
\begin{equation}
  \label{eq:purity2}
  \alpha = \sqrt{\frac
    {\left<\left(u-\left<u\right>\right)^2\right>_{s(u)}}
    {\left<\left(u-\left<u\right>\right)^2\right>_{pure}}} = 
  \frac{1}{\pi}\sqrt{-\frac{3}{2P(r_o)}
    \frac{\partial^2 P(r_o)}{\partial r^2}},
\end{equation}
which is equivalent to eqn.~\ref{eq:purity}.

In order to estimate the uncertainty in the measurement of $\alpha$,
we note that by squaring eqn.~\ref{eq:purity2} and substituting the
finite difference approximation of
\begin{equation}
  \label{eq:pdddiff}
  \frac{\partial^2 P(r_o)}{\partial r^2} 
  \simeq \frac{P\left(r_o + \Delta_r\right) + P\left(r_o - \Delta_r\right)
    - 2P\left(r_o\right)}{\Delta_r^2}
  = \frac{P^+ + P^- - 2P\left(r_o\right)}{\Delta_r^2},
\end{equation}
where $\Delta_r$ corresponds to a small frequency offset from the
measured peak power $P(r_o)$, we get
\begin{equation}
  \alpha^2 \simeq -\frac{3}{2\pi^2P(r_o)}
  \frac{P^+ + P^- - 2P\left(r_o\right)}{\Delta_r^2}.
\end{equation}
This means that
\begin{equation}
  \label{eq:purerrfact}
  \sigma_\alpha \simeq \frac{3}{4\pi^2\alpha\Delta_r^2P(r_o)}\sigma_{\Delta_P}, 
\end{equation}
where $\Delta_P = P^+ + P^- - 2P\left(r_o\right)$ and the extra factor
of $2\alpha$ comes from converting $\sigma_{\alpha^2}$ to $\sigma_\alpha$.

We can then expand the Fourier amplitude around the peak of the signal
as in eqn.~\ref{eq:powexpand2}, where we see that the amplitude
fluctuations from the sine term (which is anti-symmetric about $r_o$)
will cancel after the addition of $P(r_o+\Delta_r)$ and
$P(r_o-\Delta_r)$ in the finite difference approximation shown above.
Conversely, the fluctuations due to the cosine term (which is
symmetric about $r_o$) will add. These fluctuations can be computed by
taking the variance of the first non-constant term in the expansion of
$\cos\left(\delta_j\right)$ which is $\delta_j^2/2 =
2\pi^2\Delta_r^2(u-1/2)^2$.  The computation gives
\begin{mathletters}
  \begin{eqnarray}
    \left<\left(\delta_j^2/2 - \left<\delta_j^2/2\right>\right)^2\right> 
    & = & \left<\left(\delta_j^2/2\right)^2\right> - 
    \left<\delta_j^2/2\right>^2\\
    & = & 4\pi^4\Delta_r^4 \left\{
      \int_0^1 \left(u-\frac{1}{2}\right)^4\,\mathrm{d}u - 
      \left[\int_0^1 \left(u-\frac{1}{2}\right)^2\,\mathrm{d}u\right]^2\right\}\\
    & = & \pi^4\Delta_r^4\left[\frac{1}{20}-\left(\frac{1}{6}\right)^2\right]\\
    & = & \frac{\pi^4\Delta_r^4}{45},
  \end{eqnarray}
\end{mathletters}%
and since the variations in the $\cos\left(\phi_j\right)$ term average
to $1/2$, just like the $\sin\left(\phi_j\right)$ term did in
eqn.~\ref{eq:sine_fluct}, the standard deviation of
$\sqrt{P(r_o+\Delta_r)}$ is equal to
\begin{equation}
  \sigma_{\sqrt{P(r_o+\Delta_r)}} = 
  \sqrt{P(r_o+\Delta_r)}\frac{\pi^2\Delta_r^2}{\sqrt{90}}
  \simeq
  \pi^2\Delta_r^2\sqrt{\frac{P(r_o)}{90}}.
\end{equation}
Doubling the error when converting from amplitudes to powers and then
doubling it again due to the addition of the errors from the power
offsets in eqn.~\ref{eq:pdddiff} gives
\begin{equation}
  \sigma_{\Delta_P} = 4\pi^2\Delta_r^2\sqrt{\frac{P(r_o)}{90}}.
\end{equation}
Finally, by substituting into eqn.~\ref{eq:purerrfact}, we get
\begin{equation}
  \sigma_{\alpha} = \frac{3}{4\pi^2\alpha\Delta_r^2P(r_o)}
  4\pi^2\Delta_r^2\sqrt{\frac{P(r_o)}{90}} = \frac{1}{\alpha\sqrt{10P(r_o)}}.
\end{equation}

\section{Centroid, Purity, and Fourier Response} \label{app:response}

In order to consider the effects of centroid and purity on the Fourier
response to a sinusoidal signal as described by
eqn.~\ref{eq:cent_sigdef}, we initially assume a Fourier response equal
to eqn.~\ref{eq:cosresp} of
\begin{equation}
  \label{eq:cpdep}
  A_r = A_o\,\me^{-\mi\pi\left(r-r_o\right)}\sinc\left[\pi(r-r_o)\right]
\end{equation}
where $A_o = Na/2\;\me^{\mi \phi_o}$, $\phi_o$ is the
intrinsic phase of the signal, and $r_o$ is the true signal frequency
(in FFT bins).  This response is correct only for signals with a
square window function (i.e.\ $W(u) = 1$ from $0 \le u \le 1$ and zero
elsewhere).

From eqn.~\ref{eq:cpdep} we see that a change in Fourier frequency of
a single Fourier bin causes a change in the measured Fourier phase of
$\pi$\,radians.  This phase change is also visible from the centroid
equation for a pure signal with $\hat{C} = 1/2$, where
$\mathrm{d}\phi(r) = -\pi$ for every $\mathrm{d}r = 1$.  Rewriting the
centroid equation and integrating, we see that the Fourier phase near
the peak response goes as
\begin{equation}
  \label{eq:phase_cent}
  \phi(r) = -2\pi r \hat{C} + c.
\end{equation}
When $r=r_o$, $\phi=\phi_o$, allowing us to solve for the constant of
integration, $c = \phi_o + 2 \pi r_o \hat{C}$.  Substituting into
eqn.~\ref{eq:phase_cent}, we see that the phase of the Fourier
response is equal to
\begin{equation}
  \phi(r_o) = \phi_o - 2\pi\hat{C}(r-r_o).
\end{equation}
Therefore, for signals that have centroids different from $1/2$, the
phase change across a single Fourier bin is different from the usual
$\pi$\,radians.

The purity, as described in \S\ref{sec:sigprops} and
Appendix~\ref{app:purity}, is the effective duration of a square
window which reproduces the measured rms dispersion of the signal in
time about the centroid.  Since the Fourier response to a square
window goes as $\sinc(\pi fT)$, where $fT = r$, we can see that
replacing the window of length $T$ with one of effective duration
$\alpha T$ causes the Fourier response to go as $\sinc(\pi \alpha
fT)$.  This fact is also approximately true for more complicated
window functions as long as $\left|r-r_o\right| \ll 1$.  The Fourier
response to a windowed sinusoid is therefore
\begin{equation}
  A_r \simeq A_o\,\me^{-2\pi\mi\hat{C}(r-r_o)}\,
  \sinc\left[\pi\alpha(r-r_o)\right].
\end{equation}
Numerical simulations show that this approximation is valid for purity
values $\alpha \lesssim 1.5$.

These same ``effective duration'' arguments also apply to the shape of
the response in the \fdot\ direction of the \ffdot\ plane.  A change
in the effective duration of a signal causes a change in the \fdot\ 
response since $\dot{f}T^2 \to \dot{f}\alpha^2T^2$, or equivalently,
$\dot{r} \to \dot{r}\alpha^2$.  The results of this change can be
calculated by directly substituting $\dot{r}\alpha^2$ for \rdot\ in
eqn.~\ref{eq:ffdotresp}, as was done in Appendix~\ref{app:fdotsigma}.


\clearpage

\end{document}